\documentstyle[12pt]{article}

\font\tenrm=cmr10

 1
 1
 1

\def\ie{{\em i.e.}}
\def\eg{{\em e.g.}}

\def\beq{\begin{equation}}
\def\eeq{\end{equation}}

\catcode`\@=11 % This allows us to modify PLAIN macros.
\def\coeff#1#2{{\textstyle{#1\over #2}}}

\def\vev#1{\left\langle #1\right\rangle}
\def\lsim{\mathrel{\mathpalette\@versim<}}
\def\gsim{\mathrel{\mathpalette\@versim>}}
\def\@versim#1#2{\vcenter{\offinterlineskip
    \ialign{$\m@th#1\hfil##\hfil$\crcr#2\crcr\sim\crcr } }}
\def\etal{{\em et. al.}}
\def\JL{J. L. Lopez}
\def\DVN{D. V. Nanopoulos}

\def\r#1{$\bf#1$}
\def\rb#1{$\bf\overline{#1}$}

\def\t1{{\tilde 1}}
\def\ov{\overline}

\def\mpt{p\hskip-5.5pt/\hskip2pt}

\def\GeV{\,{\rm GeV}}
\def\TeV{\,{\rm TeV}}
\def\y{\,{\rm y}}

\def\wt{\widetilde}

\def\to{\rightarrow}
\def\pb{\,{\rm pb}}
\def\ipb{\,{\rm pb}^{-1}}
\def\NPB#1#2#3{Nucl. Phys. B {\bf#1} (19#2) #3}
\def\PLB#1#2#3{Phys. Lett. B {\bf#1} (19#2) #3}

\def\PRD#1#2#3{Phys. Rev. D {\bf#1} (19#2) #3}
\def\PRL#1#2#3{Phys. Rev. Lett. {\bf#1} (19#2) #3}
\def\PRT#1#2#3{Phys. Rep. {\bf#1} (19#2) #3}
\def\MODA#1#2#3{Mod. Phys. Lett. A {\bf#1} (19#2) #3}

\def\TAMU#1{Texas A \& M University preprint CTP-TAMU-#1}

\begin{document}
\bibliographystyle{unsrt}
% TH format
\begin{flushright}
\baselineskip=12pt
{CERN-TH.6934/93}\\
{CTP-TAMU-34/93}\\
{ACT-13/93}\\
\end{flushright}
% PPE format
%\begin{center}
%{\large EUROPEAN ORGANIZATION FOR NUCLEAR RESEARCH}
%\end{center}
%\begin{flushright}
%{CERN-PPE/93-??}\\
%{?? June, 1993}\\
%{CERN-LAA/93-??}\\
%{CERN-TH.6934/93}\\
%{CTP-TAMU-34/93}\\
%{ACT-13/93}\\
%\end{flushright}
%

\begin{center}
\vglue 0.3cm
{\Large\bf The SuperWorlds of SU(5) and SU(5) x U(1):\\}
\vspace{0.2cm}
{\Large\bf A Critical Assessment and Overview}
\footnote{To appear in the Proceedings of the International School of
Subnuclear Physics 30th Course: ``From Superstrings to the Real Superworld",
Erice, Italy, July 14-22, 1992; D. Nanopoulos and A. Zichichi Lecturers.}\\
\vglue 0.5cm
{JORGE L. LOPEZ$^{(a),(b)}$, D. V. NANOPOULOS$^{(a),(b),(c)}$, and A.
ZICHICHI$^{(d)}$\\}
\vglue 0.4cm
{\em $^{(a)}$Center for Theoretical Physics, Department of Physics, Texas A\&M
University\\}
{\em College Station, TX 77843--4242, USA\\}
{\em $^{(b)}$Astroparticle Physics Group, Houston Advanced Research Center
(HARC)\\}
{\em The Woodlands, TX 77381, USA\\}
{\em $^{(c)}$CERN, Theory Division, 1211 Geneva 23, Switzerland\\}
{\em $^{(d)}$CERN, 1211 Geneva 23, Switzerland\\}
\baselineskip=12pt

\vglue 0.4cm
{\tenrm ABSTRACT}
 \end{center}
\vglue -0.2cm
{\rightskip=3pc
 \leftskip=3pc
\xpt\baselineskip=12pt
\noindent
We present an overview of the simplest supergravity models which enforce
radiative breaking of the electroweak symmetry, namely the minimal $SU(5)$
supergravity model and the class of string-inspired/derived supergravity models
based on the flipped $SU(5)\times U(1)$ structure supplemented by a minimal set
of additional matter representations such that unification occurs at the string
scale ($\sim10^{18}\GeV$). These models can be fully parametrized in terms of
the top-quark mass, the ratio $\tan\beta=v_2/v_1$, and three supersymmetry
breaking parameters ($m_{1/2},m_0,A$). The latter are chosen in the minimal
$SU(5)$ model such that the stringent constraints from proton decay and
cosmology are satisfied. In the flipped $SU(5)$ case we consider two
string-inspired supersymmetry breaking scenaria: $SU(N,1)$ no-scale
supergravity and a dilaton-induced supersymmetry breaking scenario. Both imply
universal soft supersymmetry breaking parameters: $m_0=A=0$ and
$m_0=\coeff{1}{\sqrt{3}}m_{1/2}, A=-m_{1/2}$ respectively. We present a
comparative study of the sparticle and Higgs spectra of both flipped $SU(5)$
models and the minimal $SU(5)$ model and conclude that all can be partially
probed at the Tevatron and LEPII (and the flipped models at HERA too). In both
flipped $SU(5)$ cases there is a more constrained version which allows to
determine $\tan\beta$ in terms of $m_t,m_{\tilde g}$ and which leads to much
sharper and readily accessible experimental predictions. We also discuss the
prospects for indirect experimental detection: a non-trivial fraction of the
parameter space of the flipped $SU(5)$ models is in conflict with the present
experimental allowed range for the $b\to s\gamma$ rare decay mode, and the
one-loop electroweak radiative corrections imply the 90\% CL upper bound
$m_t\lsim175\GeV$.}
% TH format
\begin{flushleft}
\baselineskip=12pt
{CERN-TH.6934/93}\\
{CTP-TAMU-34/93}\\
{ACT-13/93}\\
June 1993
\end{flushleft}
\vfill\eject
\tableofcontents
\vfill\eject
\setcounter{page}{1}
\pagestyle{plain}

\baselineskip=14pt

\section{Introduction}
The Standard Model of electroweak and strong interactions is well established
by now. In fact, the effects of the top quark in one-loop electroweak processes
predict its mass (within $\approx20\%$) centered around $\approx145\GeV$
\cite{Rolandi}. Therefore, its expected direct experimental detection at the
Tevatron in the near future will complete the set of Standard Model predictions
for the vector and fermion sectors. The scalar sector is another  story. The
simplest electroweak symmetry breaking scenario with a single Higgs boson is
only mildly constrained experimentally, with a lower bound of $m_H\gsim60\GeV$
\cite{Ting} and no firm indirect experimental upper bound, although this
situation will change once the top quark mass is measured \cite{topup}. On the
other hand, interesting upper bounds on $m_H$ follow from various theoretical
assumptions, such as perturbative unitarity at tree- ($m_H\lsim700\GeV$)
\cite{tree} and one-loop ($m_H\lsim400\GeV$) \cite{loops} levels, and the
stability of the Higgs potential ($m_H\lsim500\GeV$) \cite{LSZ}. In practice,
with the advent of the SSC and LHC, experimental information about
the TeV scale is likely to clarify the composition of the Higgs sector.
Nevertheless, despite all these efforts the structure of the Standard Model and
its corresponding Higgs sector will remain basically unexplained.

It has therefore become customary to turn to the physics at very high energies
to search for answers to these theoretical questions. The most promising
theories of this kind contain two new ingredients: supersymmetry and
unification. Together these can explain the origin of the weak scale (\ie, the
gauge hierarchy problem) relative to the very high energy unification ($M_U$)
or Planck ($M_{Pl}$) scales  \cite{EWx}. Furthermore, this class of theories
predict a  new set of relatively light
($\lsim{\cal O}(1\TeV)$) particles consisting of partners for the Standard
Model particles but with spin offset by 1/2 unit. In fact, the new set of
particles appears ever more likely to overlap little with the mass scales of
the standard ones, thus their present unobserved status. Moreover, the
Standard Model Higgs boson will then appear as one of the new particles but
with mass close to $M_Z$, thus avoiding naturally the theoretical problems
mentioned above.

Unfortunately, the introduction of supersymmetry also increases significantly
the number of unknown parameters in the theory, mainly because this symmetry
must be softly broken at low energies. Indeed, to describe a generic
low-energy supersymmetric model (the so-called minimal supersymmetric
standard model (MSSM)) neglecting the first- and second-generation
Yukawa couplings, the CKM angles, and possible CP violating phases,
we need the following set of parameters (the values of
$\sin^2\theta_w,\alpha_3,\alpha_e,M_Z$ are taken as measured parameters):
\begin{description}
\item (a) The Yukawa ($\lambda_t,\lambda_b,\lambda_\tau$) and Higgs
mixing ($\mu$) superpotential couplings. (We can trade the Yukawa couplings
for $m_t,\tan\beta$; $m_b,m_\tau$, with $\tan\beta=v_2/v_1$ the ratio of
Higgs vacuum expectation values, and $m_b,m_\tau$ given.)
\item (b) The soft-supersymmetry breaking trilinear ($A_t,A_b,A_\tau$) and
bilinear ($B$) scalar couplings (corresponding to the superpotential couplings
in (a)).
\item (c) The soft-supersymmetry breaking left-left and right-right
entries in the squark and slepton mass matrices for the first and second
($m_{Q,U^c,D^c},\,m_{L,E^c}$), and third
($m_{Q_3,U^c_3,D^c_3},\,m_{L_3,E^c_3}$) generations.
\item (d) The soft-supersymmetry breaking gaugino masses $m_{\tilde g},
m_{\wt W},m_{\wt B}$.
\item (e) The Higgs sector parameter (at tree-level), \eg, the pseudoscalar
Higgs boson mass $m_A$.
\end{description}

\noindent The above 21 unknown parameters make any thorough analysis of this
class of models rather impractical, and have allowed in the past only limited
explorations of this parameter space. If we now add the gauge unification
constraint ($\alpha_i(M_U)=\alpha_U,\,i=1,2,3$),
the assumption of universal soft-supersymmetry breaking at a
scale $\Lambda_{susy}=M_U$, and high-energy dynamics (in the form of
renormalization group equations (RGEs) for all the parameters involved), the
set of parameters in (b) reduces to $A=A_t=A_b=A_\tau$ and $B$, those
in (c) to $m_0=m_{Q,U^c,D^c}=m_{Q_3,U^c_3,D^c_3}=m_{L,E^c}=m_{L_3,E^c_3}$,
and those in (d) to $m_{1/2}=m_{\tilde g}=m_{\wt W}=m_{\wt B}$; these
relations are valid {\it only} at the scale $\Lambda_{susy}$. The number of
parameters has been dramatically reduced down to eight.

Let us now add low-energy dynamics by demanding radiative breaking of the
electroweak symmetry. The tree-level Higgs potential is given by
\begin{eqnarray}
V_0&=&(m^2_{H_1}+\mu^2)|H_1|^2+(m^2_{H_2}+\mu^2)|H_2|^2
					+B\mu(H_1H_2+{\rm h.c.})\nonumber\\
&+&\coeff{1}{8}g^2_2(H^\dagger_2{\bf\sigma}H_2+H^\dagger_1{\bf\sigma}H_1)^2
+\coeff{1}{8}g'^2\left(|H_2|^2-|H_1|^2\right)^2,	\label{Inti}
\end{eqnarray}
where $H_1\equiv{{H^0_1\choose H^-_1}}$ and $H_2\equiv{{H^+_2\choose H^0_2}}$
are the two complex Higgs doublet fields, $g'=\sqrt{5/3}\,g_1$ and $g_2$ are
the $U(1)_Y$ and  $SU(2)_L$ gauge couplings, and $B\mu$ is taken to be real and
negative. This  potential has a minimum\footnote{The parameters in $V_0$ must
satisfy further  consistency constraints to insure that this is a true minimum
of the tree-level  Higgs potential. As discussed in Ref. \cite{GRZ},  the
one-loop effective potential satisfies most of these constraints
automatically.} if $\partial V_0/\partial\phi_i=0$, with $\phi_i$ denoting the
eight real degrees of freedom of $H_1$ and $H_2$. In particular, for
$\phi_i={\rm Re}\,H^0_i$ one obtains two constraints which allow the
determination of $\mu$ and $B$,
\begin{eqnarray}
\mu^2&=&{m^2_{H_1}-m^2_{H_2}\tan^2\beta\over\tan^2\beta-1}
-\coeff{1}{2}M^2_Z,\\
B\mu&=&-\coeff{1}{2}\sin2\beta(m^2_{H_1}+m^2_{H_2}+2\mu^2)<0,\label{Intii}
\end{eqnarray}
up to the sign of $\mu$. In these expressions, $m^2_{H_1},m^2_{H_2}$ are
soft-supersymmetry breaking masses equal to $m^2_0$ at $\Lambda_{susy}$.
Since the whole set of Higgs masses and couplings (at tree-level) follows
from $m^2_A$ (and $\tan\beta$), and one can easily show that
$m^2_A=-2B\mu/\sin2\beta$, the parameter in (e) is also determined. (This
result also holds at one-loop although the expression for $m^2_A$ is more
complicated in this case.)

The final parameter count in the class of models we consider is then just
five: $m_t,\tan\beta,m_{1/2},m_0,A$ (plus the sign of $\mu$). Note also that
$\sin^2\theta_w$ (as well as $M_U$ and $\alpha_U$) gets determined (from
$\alpha_3$ and $\alpha_e$) by the gauge unification condition. What are the
{\it a priori} expected values of $m_{1/2},m_0,A$? In principle choosing a
suitable supergravity model (\ie, a suitable hidden sector) one could have
arbitrary values for these parameters. In fact, such generic scenaria are
worthwhile studying. However, further well motivated constraints on this
sector of the theories turn out to yield quite predictive models. Below we
consider two string-inspired supersymmetry breaking scenaria where $m_0$ and
$A$ are fixed functions of $m_{1/2}$. We also consider a generic scenario which
is constrained by the proton lifetime and which results in strict restrictions
on the allowed choices for these three parameters.

In this paper we study and contrast two well motivated examples belonging
to the class of unified supersymmetric models described above, namely the
class of flipped $SU(5)$ supergravity models and the minimal $SU(5)$
supergravity model. Our purpose is to describe their features and compare
their predictions. However, we emphasize throughout that the underlying
motivation for these models is different: flipped $SU(5)$ is an archetypal
string model, whereas minimal $SU(5)$ is an archetypal traditional grand
unified model. The conceptual differences between these two frameworks are
significant, \eg:
\begin{enumerate}
\item In string models the gauge couplings of the various gauge groups
(which are typical in these constructions \cite{models}) all unify at a
calculable scale near the Planck scale.\footnote{This result holds as long as
all gauge groups are represented by level-one Kac-Moody algebras, otherwise
various factors appear in the unification relations \cite{Ginsparg}.} In fact,
the decoupling effect of the infinite tower of string massive states can be
incorporated exactly \cite{Kaplunovsky,Lacaze,ILR}. In contrast, in traditional
GUT models the unification scale is not predicted by the theory, although it
can be easily determined by running the low-energy gauge couplings until they
unify.
\item Grand unified gauge groups describing the observable sector of string
models are rare \cite{Lewellen}, if not disfavored. One can easily obtain the
Standard Model ($SU(3)\times SU(2)$) gauge groups \cite{SMOrb,SMFFF}, the
Pati-Salam ($SU(4)\times SU(2)\times SU(2)$) gauge group \cite{ALR}, and the
flipped $SU(5)$ \cite{Barr,revitalized} gauge group \cite{revamp}. Of these,
only flipped $SU(5)$ actually unifies the non-abelian gauge couplings of the
Standard Model. A proliferation of $U(1)$ factors is the norm. It is amusing to
note that flipped $SU(5)$ can be argued to be the simplest unified gauge group
\cite{revitalized,faspects}, in that $SU(5)$ does not provide for neutrino
masses, and $SO(10)$ includes flipped $SU(5)$.
\item The spectra of string models are correlated with the corresponding gauge
groups \cite{ELN} and the fermion matter representations are automatically
anomaly-free. In contrast, in traditional GUT models anomalies are actual
constraints on the possible models and the representations one can choose from
are only limited by gauge invariance. For instance, the missing partner
mechanism \cite{MPM} to effect the doublet-triplet splitting of the Higgs
pentaplets in $SU(5)$ requires the introduction of large representations which
in a string model can only occur when complicated higher-level Kac--Moody
algebra realizations of the gauge group are invoked \cite{ELN}.
\item All interactions in string models can be calculated once the model
is specified. This full specification of the effective supergravity and the
superpotential has no analog in GUT models, where all these are free
parameters.
\end{enumerate}

This paper is organized as follows.
%r In Sec. \ref{Overview} we present an overview of supersymmetric unified
%r models including gauge and Yukawa coupling unification, and radiative
%r electroweak symmetry breaking.
In Sec. \ref{Typical} we discuss the typical characteristics of string models.
In Sec. \ref{Flipped} we present in detail the flipped $SU(5)$ models which we
consider here. In Sec. \ref{Scenaria} we discuss the string-inspired
supersymmetry breaking scenaria that we explore later. In Sec. \ref{PhenoGen}
we consider the experimental predictions for all the sparticle and one-loop
corrected Higgs boson masses in the flipped $SU(5)$ models, and deduce several
simple relations among the various sparticle masses. In Sec. \ref{PhenoSp} we
repeat this analysis for special more constrained cases of the chosen
supersymmetry breaking scenaria. In Sec.~\ref{minimal} we discuss the minimal
$SU(5)$ supergravity model, including the constraints from proton decay and
the neutralino relic density. In Sec.~\ref{Direct} we discuss the prospects for
direct experimental detection of these particles at Fermilab, LEPI,II, and
HERA, while in Sec.~\ref{Indirect} we consider the  corresponding indirect
detection signatures. Finally, in Sec. \ref{Conclusions} we summarize our
conclusions.
%r In the Appendix we give a list of the RGEs used in this analysis.

%r \section{Overview of SUSY GUTs}
%r \label{Overview}
%r \subsection{Gauge coupling unification}
%r \subsection{Yukawa coupling unification}
%r \subsection{Scalar mass RGEs}
%r \subsection{Radiative electroweak symmetry breaking}
%r \subsection{Mass formulas}

\section{Typical Characteristics of String  Models}
\label{Typical}
The ultimate unification of all particles and interactions has string theory as
the best candidate. If this theory were completely understood, we would be able
to show that string theory is either inconsistent with the low-energy world
or supported by experimental data. Since our present knowledge of string theory
is at best fragmented and certainly incomplete, it is important to consider
models which incorporate as many stringy ingredients as possible. The number of
such models is expected to be large, however, the basic ingredients that such
``string models" should incorporate fall into few categories: (i) gauge group
and matter representations which unify at
a calculable model-dependent string unification scale; (ii) a hidden sector
which becomes strongly interacting at an intermediate scale and triggers
supersymmetry breaking with vanishing vacuum energy and hierarchically small
soft superpersymmetry breaking parameters; (iii) acceptable high-energy
phenomenology, \eg, gauge symmetry breaking to the Standard Model (if needed),
not-too-rapid proton decay, decoupling of intermediate-mass-scale unobserved
matter states, etc.; (iv) radiative electroweak symmetry breaking; (v)
acceptable low-energy phenomenology, \eg, reproduce the observed spectrum of
quark and lepton masses and the quark mixing angles, sparticle
and Higgs masses not in conflict with present experimental bounds, and
not-too-large neutralino cosmological relic density.

All the above are to be understood as constraints on potentially realistic
string models. Since some of the above constraints can be independently
satisfied in specific models, the real power of a string model rests in the
successful satisfaction of all these constraints within a single model.
In what follows, ``string-derived" models  refer to models which can
be derived rigorously from string, even if not all their interactions have
been determined explicitly; ``string-inspired" models are those
field-theoretical models which are believed to be in principle derivable from
string, although most likely not exactly reproducible; finally ``string" models
refer generically to both kinds, although perhaps describe more accurately
the conceptual framework these models are examples of.

String model-building is at a state of development where large numbers of
models can be constructed using various techniques (so-called formulations)
\cite{models}. Such models provide a gauge group and associated set of matter
representations, as well as all interactions in the superpotential, the
K\"ahler potential, and the gauge kinetic function. The effective string
supergravity can then be worked out and thus all the above constraints can in
principle be enforced. In practice this approach has never been followed in its
entirety: sophisticated model-building techniques exist which can produce
models satisfying constraints (i), (iii), (iv) and part of (v); detailed
studies of supersymmetry breaking triggered by gaugino condensation
have been performed for generic hidden sectors; and extensive explorations of
the soft-supersymmetry breaking parameter space satisfying constraints (iii),
(iv), and (v) have been conducted.

In searching for good string model candidates, we are faced with two kinds of
choices to be made: the choice of the gauge and matter content of the model,
and the choice of the supersymmetry breaking mechanism. Fortunately, a string
theory theorem provides significant enlightenment regarding the first choice:
models whose gauge groups are constructed from level-one Kac-Moody algebras
do not allow adjoint or higher representations in their spectra \cite{ELN}.
This implies that the traditional GUT groups ($SU(5),SO(10),E_6$) are
excluded since the GUT symmetry would remain unbroken. Exceptions to this
theorem exist if one uses the technically complicated higher-level Kac-Moody
algebras \cite{Lewellen}, but these models are beset with constraints
\cite{ELN}. If one imposes the aesthetic constraint of unification of the
Standard Model non-abelian gauge couplings, then flipped $SU(5)$
\cite{Barr,revitalized,revamp,faspects} emerges as the prime candidate, as we
shortly discuss. String models without non-abelian unification, such as the
standard-like models of Refs. \cite{SMOrb,SMFFF}
and the Pati-Salam--like model of Ref. \cite{ALR} possess nonetheless gauge
coupling unification at the string scale, even though no larger structure is
revealed past this scale. However, the degree of phenomenological success
which some of these models enjoy, usually rests on some fortuitous set of
vanishing couplings which are best understood in terms of remnants of higher
symmetries.

Besides the very economic GUT symmetry breaking mechanism in flipped $SU(5)$
\cite{Barr,revitalized} -- which allows it to be in principle derivable from
superstring theory \cite{revamp} -- perhaps one of the more interesting
motivations for considering such a unified gauge group is the natural avoidance
of potentially dangerous dimension-five proton decay operators \cite{faspects}.
In Ref. \cite{LNZI} we constructed a supergravity model based on this gauge
group, which has the additional property of unifying at a scale $M_U={\cal
O}(10^{18})\GeV$, as expected to occur in string-derived versions of this model
\cite{Lacaze}. As such, this model constitutes a blueprint for string model
builders. In fact, in Ref. \cite{LNY} one such model was derived from string
and served as inspiration for the field theory model in Ref. \cite{LNZI}.
The string unification scale should be contrasted with the naive unification
scale, $M_U={\cal O}(10^{16}\GeV)$, obtained by running the Standard Model
particles and their superpartners to very high energies. This apparent
discrepancy of two orders of magnitude \cite{EKNI,AnselmoVI} creates a {\em
gap} which needs to be bridged somehow in string models. It has been shown
\cite{price} that the simplest solution to this problem is the introduction in
the spectrum of heavy vector-like particles with Standard Model quantum
numbers. The minimal such choice \cite{sism,sismLeontaris}, a quark doublet
pair $Q,\bar Q$ and a $1/3$--charge quark singlet pair $D,\bar D$, fit snugly
inside a \r{10},\rb{10} pair of flipped $SU(5)$ representations, beyond the
usual $3\cdot({\bf10}+\ov{\bf5}+{\bf1})$ of matter and \r{10},\rb{10} of Higgs.

In this model, gauge symmetry breaking occurs due to vacuum expectation values
(vevs) of the neutral components of the \r{10},\rb{10} Higgs representations,
which develop along flat directions of the scalar potential. There are two
known ways in which these vevs (and thus the symmetry breaking scale) could
be determined:\\
\indent (i) In the conventional way, radiative corrections to the scalar
potential in the presence of soft supersymmetry breaking generate a global
minimum of the potential for values of the vevs slightly below the scale
where supersymmetry breaking effects are first felt in the observable sector
\cite{faspects}. If the latter scale is the Planck scale (in a suitable
normalization) then $M_U\sim M_{Pl}/\sqrt{8\pi}\sim 10^{18}\GeV$.\\
\indent (ii) In string-derived models a pseudo $U_A(1)$ anomaly arises as a
consequence of truncating the theory to just the massless degrees of freedom,
and adds a contribution to its $D$-term, $D_A=\sum q^A_i|\vev{\phi_i}|^2
+\epsilon$, with $\epsilon=g^2{\rm Tr\,}U_A(1)/192\pi^2\sim(10^{18}\GeV)^2$
\cite{jhreview}. To avoid a huge breaking of supersymmetry we need to demand
$D_A=0$ and therefore the fields charged under $U_A(1)$ need to get suitable
vevs. Among these one generally finds the symmetry breaking Higgs fields, and
thus $M_U\sim10^{18}\GeV$ follows.\\
\indent In general, both these mechanisms could produce somewhat lower values
of $M_U$. However, $M_U\gsim10^{16}\GeV$ is necessary to avoid too rapid proton
decay due to dimension-six operators \cite{EKNII}. In these more general cases
the $SU(5)$ and $U(1)$ gauge couplings would not unify at $M_U$ (only
$\alpha_2$ and $\alpha_3$ would), although they would eventually
``superunify" at the string scale $M_{SU}\sim10^{18}\GeV$. To simplify matters,
below we consider the simplest possible case of $M_U=M_{SU}\sim10^{18}\GeV$.
We also draw inspiration from string model-building and regard the Higgs
mixing term $\mu h\bar h$ as a result of an effective higher-order coupling
\cite{decisive,muproblem,Casasmu}, instead of as a result of a light singlet
field getting a small vev (\ie, $\lambda h\bar h\phi\to\lambda\vev{\phi}h\bar
h$) as originally considered \cite{revitalized,faspects}. An additional
contribution to $\mu$ is also generically present in supergravity models
\cite{GM,Casasmu,KL}.

The choice of supersymmetry breaking scenario is less clear. Below we show
that the phenomenologically acceptable choices basically fall in two
categories:
\begin{enumerate}
\item The no-scale ansatz \cite{LN}, which ensures the vanishing of the
(tree-level) cosmological constant even after supersymmetry breaking. This
framework also arises in the low-energy limit of superstring theory
\cite{Witten}. In a theory which contains heavy fields, the minimal no-scale
structure $SU(1,1)$ \cite{nsI} is generalized to $SU(N,1)$ \cite{nsII} which
implies that the scalar fields do not feel the supersymmetry breaking effects.
In practice this means that the universal scalar mass ($m_0$) and the universal
cubic scalar coupling ($A$) are set to zero. The sole source of supersymmetry
breaking is the universal gaugino mass ($m_{1/2}$), \ie,
\beq
m_0=0,\qquad A=0.\label{nsc}
\eeq
\item The dilaton $F$-term scenario, which also leads to universal soft
supersymmetry breaking parameters \cite{KL}
\beq
m_0=\coeff{1}{\sqrt{3}}m_{1/2},\qquad A=-m_{1/2}.\label{kl}
\eeq
\end{enumerate}
In either case, after enforcement of the
above constraints, the low-energy theory can be described in terms of just
three parameters: the top-quark mass ($m_t$), the ratio of Higgs vacuum
expectation values ($\tan\beta$), and the gluino mass ($m_{\tilde g}\propto
m_{1/2}$). Therefore, measurement of only two sparticle or Higgs masses would
determine the remaining thirty. Moreover, if the hidden sector responsible for
these patterns of soft supersymmetry breaking is specified, the gravitino mass
($m_0$) will  also be determined and the supersymmetry breaking sector of the
theory will be completely fixed.

In sum, we see basically two {\em unified} string supergravity models emerging
as good candidates for  phenomenologically acceptable string models, both of
which include a flipped $SU(5)$ observable gauge group supplemented by matter
representations in order to unify at the string scale $M_U\sim10^{18}\GeV$
\cite{price,sism,sismLeontaris}, and supersymmetry breaking is parametrized by
either of the scenaria in Eqs. (\ref{nsc},\ref{kl}).

We should remark that a real string model will include a hidden sector in
addition to the observable sector discussed in what follows. The model
presented here tacitly assumes that such hidden sector is present and that
it has suitable properties. For example, the superpotential in Eq. (\ref{W})
below, in a string model will receive contributions from cubic and higher-order
terms, with the latter generating effective observable sector couplings
once hidden sector matter condensates develop \cite{decisive}. The hidden
sector is also assumed to play a fundamental role in triggering supersymmetry
breaking via \eg, gaugino condensation. This in turn would make possible the
mechanism for gauge symmetry breaking discussed above. Probably the most
important constraint on this sector of the theory is that it should yield
one of the two supersymmetry breaking scenaria outlined above.

\section{The SU(5) x U(1) Models}
\label{Flipped}
The model we consider is a generalization of that presented in
Ref. \cite{revitalized}, and contains the following flipped $SU(5)$ fields:
\begin{enumerate}
\item three generations of quark and lepton fields $F_i,\bar f_i,l^c_i,\,
i=1,2,3$;
\item two pairs of Higgs \r{10},\rb{10} representations $H_i,\bar H_i,\,
i=1,2$;
\item one pair of ``electroweak" Higgs \r{5},\rb{5} representations
$h,\bar h$;
\item three singlet fields $\phi_{1,2,3}$.
\end{enumerate}
\noindent Under $SU(3)\times SU(2)$ the various flipped $SU(5)$ fields
decompose as follows:
\begin{eqnarray}
F_i&=&\{Q_i,d^c_i,\nu^c_i\},\quad \bar f_i=\{L_i,u^c_i\},
\quad l^c_i=e^c_i,\\
H_i&=&\{Q_{H_i},d^c_{H_i},\nu^c_{H_i}\},\quad
\bar H_i=\{Q_{\bar H_i},d^c_{\bar H_i},\nu^c_{\bar H_i}\},\\
h&=&\{H,D\},\quad \bar h=\{\bar H,\bar D\}.
\end{eqnarray}
The most general effective\footnote{To be understood in the string context as
arising from cubic and higher order terms \cite{KLN,decisive}.}
superpotential consistent with $SU(5)\times U(1)$ symmetry is given by
\begin{eqnarray}
W&=&\lambda^{ij}_1 F_iF_jh+\lambda^{ij}_2 F_i\bar f_j \bar h
+\lambda^{ij}_3 \bar f_il^c_j h +\mu h\bar h
+\lambda^{ij}_4 H_iH_jh+\lambda^{ij}_5\bar H_i\bar H_j\bar h\nonumber\\
&+&\lambda^{ij}_{1'}H_iF_jh+\lambda^{ij}_{2'}H_i\bar f_j\bar h
+\lambda^{ijk}_6 F_i\bar H_j\phi_k+w^{ij}H_i\bar H_j+\mu^{ij}\phi_i\phi_j.
\label{W}
\end{eqnarray}
Symmetry breaking is effected by non-zero vevs $\vev{\nu^c_{H_i}}=V_i$,
$\vev{\nu^c_{\bar H_i}}=\bar V_i$, such that
$V^2_1+V^2_2=\bar V^2_1+\bar V^2_2$.
\subsection{Higgs doublet and triplet mass matrices}
The Higgs doublet mass matrix receives contributions from
$\mu h\bar h\to \mu H\bar H$ and $\lambda^{ij}_{2'}H_i\bar f_j\bar h\to
\lambda^{ij}_{2'}V_i L_j\bar H$. The resulting matrix is
\beq
{\cal M}_2=\bordermatrix{&\bar H\cr H&\mu\cr
L_1&\lambda^{i1}_{2'}V_i\cr
L_2&\lambda^{i2}_{2'}V_i\cr
L_3&\lambda^{i3}_{2'}V_i\cr}.
\eeq
To avoid fine-tunings of the $\lambda^{ij}_{2'}$ couplings we must demand
$\lambda^{ij}_{2'}\equiv0$, so that $\bar H$ remains light.

The Higgs triplet matrix receives several contributions:
$\mu h\bar h\to\mu D\bar D$;
$\lambda^{ij}_{1'}H_iF_jh\to\lambda^{ij}_{1'}V_id^c_jD$;
$\lambda^{ij}_4 H_iH_jh\to\lambda^{ij}_4V_i d^c_{H_j}D$;
$\lambda^{ij}_5 \bar H_i\bar H_j\bar h
\to\lambda^{ij}_5\bar V_i d^c_{\bar H_j}\bar D$;
$w^{ij}d^c_{H_i}d^c_{\bar H_j}$. The resulting matrix is\footnote{The zero
entries in ${\cal M}_3$ result from the assumption $\vev{\phi_k}=0$ in
$\lambda_6^{ijk}F_i\bar H_j\phi_k$.}
\beq
{\cal M}_3=\bordermatrix{
&\bar D&d^c_{H_1}&d^c_{H_2}&d^c_1&d^c_2&d^c_3\cr
D&\mu&\lambda^{i1}_4V_i&\lambda^{i2}_4V_i&\lambda^{i1}_{1'}V_i
&\lambda^{i2}_{1'}V_i&\lambda^{i3}_{1'}V_i\cr
d^c_{\bar H_1}&\lambda^{i1}_5\bar V_i&w_{11}&w_{12}&0&0&0\cr
d^c_{\bar H_2}&\lambda^{i2}_5\bar V_i&w_{21}&w_{22}&0&0&0\cr}.\label{IIb}
\eeq
Clearly three linear combinations of $\{\bar D,d^c_{H_{1,2}},d^c_{1,2,3}\}$
will remain light. In fact, such a general situation will induce a mixing
in the down-type Yukawa matrix $\lambda^{ij}_1 F_iF_jh\to\lambda^{ij}_1Q_i
d^c_jH$, since the $d^c_j$ will need to be re-expressed in terms of these
mixed light eigenstates.\footnote{Note that this mixing is on top of any
structure that $\lambda^{ij}_1$ may have, and is the only source of mixing in
the typical string model-building case of a diagonal $\lambda_2$ matrix.} This
low-energy quark-mixing mechanism is an explicit realization of the general
extra-vector-abeyance (EVA) mechanism of Ref. \cite{EVA}. As a first
approximation though, in what follows we will set $\lambda^{ij}_{1'}=0$, so
that the light eigenstates are $d^c_{1,2,3}$.
\subsection{Neutrino see-saw matrix}
The see-saw neutrino matrix receives contributions from:
$\lambda^{ij}_2F_i\bar f_j\bar h\to m^{ij}_u\nu^c_i\nu_j$;
$\lambda^{ijk}_6 F_i\bar H_j\phi_k\to\lambda^{ijk}_6\bar V_j\nu^c_i\phi_k$;
$\mu^{ij}\phi_i\phi_j$. The resulting matrix is\footnote{We neglect a possible
higher-order contribution which could produce a non-vanishing $\nu^c_i\nu^c_j$
entry \cite{chorus}.}
\beq
{\cal M}_\nu=\bordermatrix{&\nu_j&\nu^c_j&\phi_j\cr
\nu_i&0&m^{ji}_u&0\cr
\nu^c_i&m^{ij}_u&0&\lambda^{ikj}_6\bar V_k\cr
\phi_i&0&\lambda^{jki}_6\bar V_k&\mu^{ij}\cr}.
\eeq
\subsection{Numerical scenario}
To simplify the discussion we will assume, besides\footnote{In Ref.
\cite{revitalized} the discrete symmetry $H_1\to -H_1$ was imposed so that
these couplings automatically vanish when $H_2,\bar H_2$ are not present. This
symmetry (generalized to $H_i\to -H_i$) is not needed here since it would
imply $w^{ij}\equiv0$, which is shown below to be disastrous for gauge
coupling unification.} $\lambda^{ij}_{1'}=\lambda^{ij}_{2'}\equiv0$, that
\begin{eqnarray}
\lambda^{ij}_4&=&\delta^{ij}\lambda^{(i)}_4,\quad
\lambda^{ij}_5=\delta^{ij}\lambda^{(i)}_5,\quad
\lambda^{ijk}_6=\delta^{ij}\delta^{ik}\lambda^{(i)}_6,\\
\mu^{ij}&=&\delta^{ij}\mu_i,\quad w^{ij}=\delta^{ij}w_i.
\end{eqnarray}
These choices are likely to be realized in string versions of this model
and will not alter our conclusions below. In this case the Higgs triplet mass
matrix reduces to
\beq
{\cal M}_3=\bordermatrix
{&\bar D&d^c_{H_1}&d^c_{H_2}\cr
D&\mu&\lambda^{(1)}_4V_1&\lambda^{(2)}_4V_2\cr
d^c_{\bar H_1}&\lambda^{(1)}_5\bar V_1&w_1&0\cr
d^c_{\bar H_2}&\lambda^{(2)}_5\bar V_2&0&w_2\cr}.\label{III}
\eeq
Regarding the $(3,2)$ states, the scalars get either eaten by the $X,Y$ $SU(5)$
heavy gauge bosons or become heavy Higgs bosons, whereas the fermions interact
with the $\wt X,\wt Y$ gauginos through the following mass matrix \cite{LNY}
\beq
{\cal M}_{(3,2)}=\bordermatrix
{&Q_{\bar H_1}&Q_{\bar H_2}&\wt Y\cr
Q_{H_1}&w_1&0&g_5V_1\cr
Q_{H_2}&0&w_2&g_5V_2\cr
\wt X&g_5\bar V_1&g_5\bar V_2&0\cr}.
\eeq
The lightest eigenvalues of these two matrices (denoted generally by $d^c_H$
and $Q_H$ respectively) constitute the new relatively light particles in the
spectrum, which are hereafter referred to as the ``{\em gap}" particles since
with suitable masses they bridge the gap between unification masses at
$10^{16}\GeV$ and $10^{18}\GeV$.

Guided by the phenomenological requirement on the gap particle masses, \ie,
$M_{Q_H}\gg M_{d^c_H}$ \cite{sism}, we consider the following explicit
numerical scenario
\beq
\lambda^{(2)}_4=\lambda^{(2)}_5=0,\quad
V_1,\bar V_1,V_2,\bar V_2\sim V\gg w_1\gg w_2\gg\mu,\label{V}
\eeq
which would need to be reproduced in a viable string-derived model. From Eq.
(\ref{III}) we then get $M_{d^c_{H_2}}=M_{d^c_{\bar H_2}}=w_2$, and all
other mass eigenstates $\sim V$. Furthermore, ${\cal M}_{(3,2)}$ has a
characteristic polynomial $\lambda^3-\lambda^2(w_1+w_2)-\lambda(2V^2-w_1w_2)
+(w_1+w_2)V^2=0$, which has two roots of ${\cal O}(V)$ and one root of
${\cal O}(w_1)$. The latter corresponds to $\sim(Q_{H_1}-Q_{H_2})$ and
$\sim(Q_{\bar H_1}-Q_{\bar H_2})$. In sum then, the gap particles have masses
$M_{Q_H}\sim w_1$ and $M_{d^c_H}\sim w_2$, whereas all other heavy particles
have masses $\sim V$.

The see-saw matrix reduces to
\beq
{\cal M}_\nu=\bordermatrix{&\nu_i&\nu^c_i&\phi_i\cr
\nu_i&0&m^i_u&0\cr
\nu^c_i&m^i_u&0&\lambda^{(i)}\bar V_i\cr
\phi_i&0&\lambda^{(i)}\bar V_i&\mu^i\cr},
\eeq
for each generation. The physics of this see-saw matrix has been
discussed in Ref. \cite{chorus} and more generally in Ref. \cite{ELNO},
where it was shown to lead to an interesting amount of hot dark matter
($\nu_\tau$) and an MSW-effect ($\nu_e,\nu_\mu$) compatible with all solar
neutrino data. Moreover, the out-of-equilibrium decays of the $\nu^c$
``flipped neutrino" fields in the early Universe induce a lepton number
asymmetry which is later processed into a baryon number asymmetry by
non-perturbative electroweak processes \cite{ENO,ELNO}. All these phenomena can
occur in the same region of parameter space.
\subsection{Proton decay}
The dimension-six operators mediating proton decay in this model are highly
suppressed due to the large mass of the $X,Y$ gauge bosons
($\sim M_U=10^{18}\GeV$). Higgsino mediated dimension-five operators exist
and are naturally suppressed in the minimal model of Ref. \cite{revitalized}.
The reason for this is that the Higgs triplet mixing term
$\mu h\bar h\to \mu D\bar D$ is small ($\mu\sim M_Z$), whereas the Higgs
triplet mass eigenstates obtained from Eq. (\ref{IIb}) by just keeping the
$2\times2$ submatrix in the upper left-hand corner, are always very heavy
($\sim V$). The dimension-five mediated operators are then proportional to
$\mu/V^2$ and thus the rate is suppressed by a factor or $(\mu/V)^2\ll1$
relative to the unsuppressed case found in the standard $SU(5)$ model.

In the generalized model presented here, the Higgs triplet mixing term is
still $\mu D\bar D$. However, the exchanged mass eigenstates are not
necessarily all very heavy. In fact, above we have demanded the existence of a
relatively light ($\sim w_1$) Higgs triplet state ($d^c_H$). In this case
the operators are proportional to $\mu\alpha_i\bar\alpha_i/{\cal M}^2_i$, where
${\cal M}_i$ is the mass of the $i$-th exchanged eigenstate and
$\alpha_i,\bar\alpha_i$ are its $D,\bar D$ admixtures. In the scenario
described above, the relatively light eigenstates ($d^c_{H_2},d^c_{\bar H_2}$)
contain no $D,\bar D$ admixtures, and the operator will again be
$\propto\mu/V^2$.

Note however that if conditions (\ref{V}) (or some analogous suitability
requirement) are not satisfied, then diagonalization of ${\cal M}_3$ in Eq.
(\ref{III}) may re-introduce a sizeable dimension-five mediated proton decay
rate, depending on the value of the $\alpha_i,\bar\alpha_i$ coefficients. To be
safe one should demand \cite{ANpd,HMY,LNPZ}
\beq
{\mu\alpha_i\bar\alpha_i\over {\cal M}^2_i}\lsim{1\over 10^{17}\GeV}.
\eeq
For the higher values of $M_{d^c_H}$ in Table \ref{Table1} (see below), this
constraint can be satisfied for not necessarily small values of
$\alpha_i,\bar\alpha_i$.
\begin{table}[p]
\hrule
\caption{The value of the gap particle masses and the unified coupling for
$\alpha_3(M_Z)=0.118\pm0.008$. We have taken $M_U=10^{18}\GeV$,
$\sin^2\theta_w=0.233$, and $\alpha^{-1}_e=127.9$.}
\label{Table1}
\begin{center}
\begin{tabular}{|c|c|c|c|}\hline
$\alpha_3(M_Z)$&$M_{d^c_H}\,(\GeV)$&$M_{Q_H}\,(\GeV)$&$\alpha(M_U)$\\ \hline
$0.110$&$4.9\times10^4\GeV$&$2.2\times10^{12}\GeV$&$0.0565$\\
$0.118$&$4.5\times10^6\GeV$&$4.1\times10^{12}\GeV$&$0.0555$\\
$0.126$&$2.3\times10^8\GeV$&$7.3\times10^{12}\GeV$&$0.0547$\\ \hline
\end{tabular}
\end{center}
\hrule
\end{table}

\begin{figure}[p]
\vspace{4in}
%\special{psfile=proc_erice1.ps angle=90 hscale=70 vscale=70 hoffset=500}
\vspace{-0.7in}
\caption{\baselineskip=12pt
The running of the gauge couplings in the flipped $SU(5)$ model for
$\alpha_3(M_Z)=0.118$ (solid lines). The gap particle masses have been derived
using the gauge coupling RGEs to achieve unification at $M_U=10^{18}\GeV$. The
case with no gap particles (dotted lines) is also shown; here
$M_U\approx10^{16}\GeV$.}
\label{Figure1}
\end{figure}

\subsection{Gauge coupling unification}
Since we have chosen $V\sim M_U=M_{SU}=10^{18}\GeV$, this means that the
Standard Model gauge couplings should unify at the scale $M_U$. However, their
running will be modified due to the presence of the gap particles. Note that
the underlying flipped $SU(5)$ symmetry, even though not evident in this
respect, is nevertheless essential in the above discussion. The masses $M_Q$
and $M_{d^c_H}$ can then be determined, as follows \cite{sism}
\begin{eqnarray}
\ln{M_{Q_H}\over m_Z}&=&\pi\left({1\over2\alpha_e}-{1\over3\alpha_3}
-{\sin^2\theta_w-0.0029\over\alpha_e}\right)-2\ln{M_U\over m_Z}-0.63,\\
\ln{M_{d^c_H}\over m_Z}&=&\pi\left({1\over2\alpha_e}-{7\over3\alpha_3}
+{\sin^2\theta_w-0.0029\over\alpha_e}\right)
			-6\ln{M_U\over m_Z}-1.47,\label{VIb}
\end{eqnarray}
where $\alpha_e$, $\alpha_3$ and $\sin^2\theta_w$ are all measured at $M_Z$.
This is a one-loop determination (the constants account for the dominant
two-loop corrections) which neglects all low- and high-energy threshold
effects,\footnote{Here we assume a common supersymmetric threshold at $M_Z$.
In fact, the supersymmetric threshold and the $d^c_H$ mass are anticorrelated.
See Ref. \cite{sism} for a discussion.} but is quite adequate for our present
purposes. As shown in Table \ref{Table1} (and Eq. (\ref{VIb})) the $d^c_H$ mass
depends most sensitively on $\alpha_3(M_Z)=0.118\pm0.008$ \cite{Bethke},
whereas the $Q_H$ mass and the unified coupling are rather insensitive to it.
The unification of the gauge couplings is shown in Fig. \ref{Figure1} (solid
lines) for the central value of $\alpha_3(M_Z)$. This figure also shows the
case of no gap particles (dotted lines), for which $M_U\approx10^{16}\GeV$.

\section{String-inspired Supersymmetry Breaking Scenaria}
\label{Scenaria}
Supersymmetry breaking in string models can generally be triggered in a
phenomenologically acceptable way by non-zero $F$-terms for: (a) any of the
moduli fields of the string model ($\vev{F_M}$) \cite{FM}, (b) the dilaton
field ($\vev{F_D}$) \cite{KL}, or (c) the hidden matter fields ($\vev{F_H}$)
\cite{AELN}. It has been recently noted  \cite{KL} that much model-independent
information can be obtained about the structure of the soft supersymmetry
breaking parameters in generic string supergravity models if one neglects the
third possibility ($\vev{F_H}=0$) and assumes that either: (i)
$\vev{F_M}\gg\vev{F_D}$, or (ii) $\vev{F_D}\gg\vev{F_M}$.

In case (i) the scalar masses are generally not universal, \ie, $m_i=f_im_0$
where $m_0$ is the gravitino mass and $f_i$ are calculable constants,
and therefore large flavor-changing-neutral-currents (FCNCs) \cite{EN} are
potentially dangerous \cite{IL}. The gaugino masses arise from the one-loop
contribution to the gauge kinetic function and are thus suppressed
($m_{1/2}\sim(\alpha/4\pi)m_0$) \cite{IL,Casas,KL}. The experimental
constraints on the gaugino masses then force the squark and slepton masses
into the TeV range \cite{Casas}. It is interesting to note that this
supersymmetry breaking scenario is not unlike that required for the minimal
$SU(5)$ supergravity model in order to have the dimension-five proton decay
operators under control \cite{ANpd,LNP,LNPZ}, which requires
$m_{1/2}/m_0\lsim\coeff{1}{3}$, as discussed in Sec.~\ref{su5pdecay}. This
constraint entails potential cosmological troubles: the neutralino relic
density is large and one needs to tune the parameters to have the neutralino
mass be very near the Higgs and $Z$ resonances
\cite{troubles,LNP,LNPZ,ANcosm,poles} (see Sec.~\ref{su5dm}). Clearly, such
cosmological constraints are going to be exacerbated in the case (i) scenario
($m_{1/2}/m_0\ll1$) and will likely require real fine-tuning of the model
parameters.

An important exception to case (i) occurs if $f_i\equiv0$ and all scalar masses
at the unification scale vanish ($\vev{F_M}_{m_0=0}$), as is the case in
unified no-scale supergravity models \cite{LN}, where the minimal no-scale
structure $SU(1,1)$ \cite{nsI} is generalized to $SU(N,1)$ \cite{nsII} in the
presence of the heavy GUT fields. This special case automatically restores the
much needed universality of scalar masses, and in the context of
no-scale models also entails $A=0$, see Eq. (\ref{nsc}). A special case of this
scenario occurs when the bilinear soft-supersymmetry breaking mass parameter
$B(M_U)$ is also required to vanish. With the additional ingredient of a
flipped $SU(5)$ gauge group, all the above problems are naturally avoided
\cite{LNZI}, and interesting predictions for direct
\cite{LNWZ,LNPWZh,LNPWZ,hera} and indirect \cite{bsgamma,ewcorr,NT}
experimental detection follow.

If supersymmetry breaking is triggered by $\vev{F_D}$ (case (ii)), one obtains
{\em universal} soft-supersymmetry gaugino and scalar masses and trilinear
interactions \cite{KL} and the soft-supersymmetry breaking parameters
in Eq. (\ref{kl}) result. As well, there is a special more constrained case
where $B(M_U)=2m_0={2\over\sqrt{3}}m_{1/2}$ is also required, if one demands
that the $\mu$ parameter receive contributions solely from supergravity
\cite{KL}. With the complement of a flipped $SU(5)$ structure, this model has
also been seen to avoid all the difficulties of the generic $\vev{F_M}$
scenario \cite{LNZII}. This supersymmetry breaking scenario has been studied
recently also in the context of the minimal supersymmetric Standard Model
(MSSM) in Ref. \cite{BLM}.

More generally, presumably it should be possible to find suitable hidden
sectors where an arbitrary choice of $m_0,m_{1/2},A$ is realized, keeping
the vacuum energy at zero. This is the attitude taken below when studying the
minimal $SU(5)$ supergravity model, where the proton decay constraint imposes
severe restrictions on the allowed choices of these three parameters. In the
context of string-inspired models (\ie, the flipped $SU(5)$ case), in what
follows we restrict ourselves to the two supersymmetry breaking scenaria in
Eqs. (\ref{nsc},\ref{kl}) and their special cases ($B(M_U)=0$ and
$B(M_U)=2m_0$, respectively).

\section{Flipped Phenomenology: General Case}
\label{PhenoGen}
The procedure to extract the low-energy predictions of the models outlined
above is rather standard by now (see \eg, Ref. \cite{aspects}): (a) the
bottom-quark and tau-lepton masses, together with the input values of $m_t$ and
$\tan\beta$ are used to determine the respective Yukawa couplings at the
electroweak scale; (b) the gauge and Yukawa couplings are then run up to the
unification scale $M_U=10^{18}\GeV$ taking into account the extra vector-like
quark doublet ($\sim10^{12}\GeV$) and singlet ($\sim10^6\GeV$) introduced above
\cite{sism,LNZI}; (c) at the unification scale the soft-supersymmetry breaking
parameters are introduced (according to Eqs. (\ref{nsc},\ref{kl})) and the
scalar masses are then run down to the electroweak scale; (d) radiative
electroweak symmetry breaking is enforced by minimizing the one-loop effective
potential which depends on the whole mass spectrum, and the values of the Higgs
mixing term $|\mu|$ and the bilinear soft-supersymmetry breaking parameter $B$
are determined from the minimization conditions; (e) all known phenomenological
constraints on the sparticle and Higgs masses are applied (most importantly the
LEP lower bounds on the chargino and Higgs masses), including the cosmological
requirement of not-too-large neutralino relic density.
\subsection{Mass ranges}
We have scanned the parameter space for $m_t=130,150,170\GeV$,
$\tan\beta=2\to50$ and $m_{1/2}=50\to500\GeV$. Imposing the constraint
$m_{\tilde g,\tilde q}<1\TeV$ we find
\begin{eqnarray}
&\vev{F_M}_{m_0=0}:\qquad	&m_{1/2}<475\GeV,\quad \tan\beta\lsim32,\\
&\vev{F_D}:\qquad	&m_{1/2}<465\GeV,\quad \tan\beta\lsim46.
\end{eqnarray}
These restrictions on $m_{1/2}$ cut off the growth of most of the sparticle and
Higgs masses at $\approx1\TeV$. However, the sleptons, the lightest Higgs, the
two lightest neutralinos, and the lightest chargino are cut off at a much lower
mass, as follows\footnote{In this class of supergravity models the three
sneutrinos ($\tilde\nu$) are degenerate in mass. Also, $m_{\tilde
\mu_L}=m_{\tilde e_L}$ and $m_{\tilde\mu_R}=m_{\tilde e_R}$.}
\begin{eqnarray}
&\vev{F_M}_{m_0=0}:&\left\{
	\begin{array}{l}
	m_{\tilde e_R}<190\GeV,\quad m_{\tilde e_L}<305\GeV,
				\quad m_{\tilde\nu}<295\GeV\\
	m_{\tilde\tau_1}<185\GeV,\quad m_{\tilde\tau_2}<315\GeV\\
	m_h<125\GeV\\
	m_{\chi^0_1}<145\GeV,\quad m_{\chi^0_2}<290\GeV,
				\quad m_{\chi^\pm_1}<290\GeV
	\end{array}
		\right.\\
&\vev{F_D}:&\left\{
	\begin{array}{l}
	m_{\tilde e_R}<325\GeV,\quad m_{\tilde e_L}<400\GeV,
				\quad m_{\tilde\nu}<400\GeV\\
	m_{\tilde\tau_1}<325\GeV,\quad m_{\tilde\tau_2}<400\GeV\\
	m_h<125\GeV\\
	m_{\chi^0_1}<145\GeV,\quad m_{\chi^0_2}<285\GeV,
				\quad m_{\chi^\pm_1}<285\GeV
	\end{array}
		\right.
\end{eqnarray}
It is interesting to note that due to the various constraints on the model,
the gluino and (average) squark masses are bounded from below,
\beq
\vev{F_M}_{m_0=0}:\left\{
	\begin{array}{l}
	m_{\tilde g}\gsim245\,(260)\GeV\\
	m_{\tilde q}\gsim240\,(250)\GeV
	\end{array}
		\right.
\qquad
\vev{F_D}:\left\{
	\begin{array}{l}
	m_{\tilde g}\gsim195\,(235)\GeV\\
	m_{\tilde q}\gsim195\,(235)\GeV
	\end{array}
		\right.		\label{gmin}
\eeq
for $\mu>0(\mu<0)$. Relaxing the above conditions on $m_{1/2}$ simply allows
all sparticle masses to grow further proportional to $m_{\tilde g}$.

\begin{table}
\hrule
\caption{
The value of the $c_i$ coefficients appearing in  Eq.~(28), the ratio
$c_{\tilde g}=m_{\tilde g}/m_{1/2}$, and the average squark coefficient
$\bar c_{\tilde q}$, for $\alpha_3(M_Z)=0.118\pm0.008$. Also shown are the
$a_i,b_i$ coefficients for the central value of $\alpha_3(M_Z)$ and both
supersymmetry breaking scenaria ($M:\vev{F_M}_{m_0=0}$, $D:\vev{F_D}$).
 The results apply as well to the second-generation squark and slepton masses.}
\label{Table2}
\begin{center}
\begin{tabular}{|c|c|c|c|}\hline
$i$&$c_i\,(0.110)$&$c_i\,(0.118)$&$c_i\,(0.126)$\\ \hline
$\tilde\nu,\tilde e_L$&$0.406$&$0.409$&$0.413$\\
$\tilde e_R$&$0.153$&$0.153$&$0.153$\\
$\tilde u_L,\tilde d_L$&$3.98$&$4.41$&$4.97$\\
$\tilde u_R$&$3.68$&$4.11$&$4.66$\\
$\tilde d_R$&$3.63$&$4.06$&$4.61$\\
$c_{\tilde g}$&$1.95$&$2.12$&$2.30$\\
$\bar c_{\tilde q}$&$3.82$&$4.07$&$4.80$\\ \hline
\end{tabular}
\begin{tabular}{|c|c|c|c|c|}\hline
$i$&$a_i(M)$&$b_i(M)$&$a_i(D)$&$b_i(D)$\\ \hline
$\tilde e_L$&$0.302$&$+1.115$&$0.406$&$+0.616$\\
$\tilde e_R$&$0.185$&$+2.602$&$0.329$&$+0.818$\\
$\tilde\nu$&$0.302$&$-2.089$&$0.406$&$-1.153$\\
$\tilde u_L$&$0.991$&$-0.118$&$1.027$&$-0.110$\\
$\tilde u_R$&$0.956$&$-0.016$&$0.994$&$-0.015$\\
$\tilde d_L$&$0.991$&$+0.164$&$1.027$&$+0.152$\\
$\tilde d_R$&$0.950$&$-0.033$&$0.989$&$-0.030$\\ \hline
\end{tabular}
\end{center}
\hrule
\end{table}

\subsection{Mass relations}
The neutralino and chargino masses show a correlation observed before in
this class of models \cite{ANc,LNZI}, namely
\beq
m_{\chi^0_1}\approx \coeff{1}{2}m_{\chi^0_2},\qquad
m_{\chi^0_2}\approx m_{\chi^\pm_1}\approx M_2=(\alpha_2/\alpha_3)m_{\tilde g}
\approx0.28m_{\tilde g}.\label{neuchar}
\eeq
This is because throughout the parameter space $|\mu|$ is generally much larger
than $M_W$ and $|\mu|>M_2$. In practice we find $m_{\chi^0_2}\approx
m_{\chi^\pm_1}$ to be satisfied quite accurately, whereas
$m_{\chi^0_1}\approx{1\over2}m_{\chi^0_2}$ is only qualitatively satisfied,
although the agreement is better in the $\vev{F_D}$ case. In fact, these two
mass relations are much more reliable than the one that links them to
$m_{\tilde g}$. The heavier neutralino ($\chi^0_{3,4}$) and chargino
($\chi^\pm_2$) masses are determined by the value of $|\mu|$; they all approach
this limit for large enough $|\mu|$. More precisely, $m_{\chi^0_3}$ approaches
$|\mu|$ sooner than $m_{\chi^0_4}$ does. On the other hand, $m_{\chi^0_4}$
approaches $m_{\chi^\pm_2}$ rather quickly.

The first- and second-generation squark and slepton masses can be determined
analytically
\beq
\wt m_i=\left[m^2_{1/2}(c_i+\xi^2_0)-d_i{\tan^2\beta-1\over\tan^2\beta+1}
M^2_W\right]^{1/2}=a_i m_{\tilde g}\left[1+b_i\left({150\over m_{\tilde
g}}\right)^2{\tan^2\beta-1\over\tan^2\beta+1}\right]^{1/2},\label{masses}
\eeq
where $d_i=(T_{3i}-Q)\tan^2\theta_w+T_{3i}$ (\eg, $d_{\tilde u_L}={1\over2}
-{1\over6}\tan^2\theta_w$, $d_{\tilde e_R}=-\tan^2\theta_w$), and
$\xi_0=m_0/m_{1/2}=0,\coeff{1}{\sqrt{3}}$. The coefficients $c_i$ can be
calculated numerically in terms of the low-energy gauge couplings, and are
given in  Table \ref{Table2}\footnote{These are renormalized at the scale
$M_Z$. In a more accurate treatment, the $c_i$ would be renormalized at the
physical sparticle mass scale, leading to second order shifts on the sparticle
masses.} for $\alpha_3(M_Z)=0.118\pm0.008$. In the table we also give
$c_{\tilde g}=m_{\tilde g}/m_{1/2}$. Note that these values are smaller than
what is obtained in the minimal $SU(5)$ supergravity model (where $c_{\tilde
g}=2.90$ for $\alpha_3(M_Z)=0.118$) and therefore the numerical relations
between the gluino mass and the neutralino masses are different in that model.
In the table we also show the resulting values for $a_i,b_i$ for the central
value of $\alpha_3(M_Z)$. Note that the apparently larger $\tan\beta$
dependence in the $\vev{F_M}_{m_0=0}$ case (\ie, $|b_i(M)|>|b_i(D)|$) is
actually compensated by a larger minimum value of $m_{\tilde g}$ in this case
(see Eq. (\ref{gmin})).

The ``average" squark mass, $m_{\tilde q}\equiv{1\over8}(m_{\tilde
u_L}+m_{\tilde u_R}+m_{\tilde d_L}+m_{\tilde d_R}+m_{\tilde c_L}+m_{\tilde c_R}
+m_{\tilde s_L}+m_{\tilde s_R})
=(m_{\tilde g}/c_{\tilde q})\sqrt{\bar c_{\tilde q}+\xi^2_0}$, with $\bar
c_{\tilde q}$ given in Table \ref{Table2}, is determined to be
\beq
m_{\tilde q}=\left\{	\begin{array}{ll}
			(1.00,0.95,0.95) m_{\tilde g},&\quad\vev{F_M}_{m_0=0}\\
			(1.05,0.99,0.98) m_{\tilde g},&\quad\vev{F_D}
			\end{array}
		\right.
\eeq
for $\alpha_3(M_Z)=0.110,0.118,0.126$ (the dependence on $\tan\beta$ is small).
The squark splitting around the average is $\approx2\%$.

These masses are plotted in Fig. \ref{Figure2}. The thickness and straightness
of the lines shows the small $\tan\beta$ dependence, except for $\tilde\nu$.
The results do not depend on the sign of $\mu$, except to the extent that some
points in parameter space are not allowed for both signs of $\mu$: the $\mu<0$
lines start-off at larger mass values. Note that
\beq
\vev{F_M}_{m_0=0}:\left\{
	\begin{array}{l}
	m_{\tilde e_R}\approx0.18m_{\tilde g}\\
	m_{\tilde e_L}\approx0.30m_{\tilde g}\\
	m_{\tilde e_R}/m_{\tilde e_L}\approx0.61
	\end{array}
		\right.
\qquad
\vev{F_D}:\left\{
	\begin{array}{l}
	m_{\tilde e_R}\approx0.33m_{\tilde g}\\
	m_{\tilde e_L}\approx0.41m_{\tilde g}\\
	m_{\tilde e_R}/m_{\tilde e_L}\approx0.81
	\end{array}
		\right.
\eeq

\begin{figure}[p]
\vspace{4.7in}
%\special{psfile=proc_erice2a.ps angle=90 hscale=60 vscale=60 hoffset=440}
\vspace{3.8in}
%\special{psfile=proc_erice2b.ps angle=90 hscale=60 vscale=60 hoffset=440}
\vspace{-0.7in}
\caption{\baselineskip=12pt
The first-generation squark and slepton masses as a function of
the gluino mass, for both signs of $\mu$, $m_t=150\GeV$, and both supersymmetry
breaking scenaria under consideration. The same values apply to the second
generation. The thickness of the lines and their deviation from linearity are
because of the small $\tan\beta$ dependence.}
\label{Figure2}
\end{figure}

The third generation squark and slepton masses cannot be determined
analytically. In Fig. \ref{Figure3} we show
$\tilde\tau_{1,2},\tilde b_{1,2},\tilde t_{1,2}$ for the choice $m_t=150\GeV$.
The variability on the $\tilde\tau_{1,2}$ and $\tilde b_{1,2}$ masses
is due to the $\tan\beta$-dependence in the off-diagonal element of the
corresponding $2\times2$ mass matrices ($\propto
m_{\tau,b}(A_{\tau,b}+\mu\tan\beta)$). The off-diagonal element in the
stop-squark mass matrix ($\propto m_t(A_t+\mu/\tan\beta)$) is
rather insensitive to $\tan\beta$ but still effects a large $\tilde t_1-\tilde
t_2$ mass splitting because of the significant $A_t$ contribution. Note
that both these effects are more pronounced for the $\vev{F_D}$ case since
there $|A_{t,b,\tau}|$ are larger than in the $\vev{F_M}_{m_0=0}$ case.
The lowest values of the $\tilde t_1$ mass go up with $m_t$ and can be as low
as
\beq
m_{\tilde t_1}\gsim\left\{	\begin{array}{ll}
		160,170,190\,(155,150,170)\GeV;&\quad\vev{F_M}_{m_0=0}\\
			88,112,150\,(92,106,150)\GeV;&\quad\vev{F_D}
			\end{array}
		\right.
\eeq
for $m_t=130,150,170\GeV$ and $\mu>0\,(\mu<0)$.

\begin{figure}[p]
\vspace{4.3in}
%\special{psfile=proc_erice3a.ps angle=90 hscale=60 vscale=55 hoffset=440}
\vspace{3.8in}
%\special{psfile=proc_erice3b.ps angle=90 hscale=60 vscale=55 hoffset=440}
\vspace{-0.5in}
\caption{\baselineskip=12pt
The $\tilde\tau_{1,2}$, $\tilde b_{1,2}$, and $\tilde t_{1,2}$ masses
versus the gluino mass for both signs of $\mu$, $m_t=150\GeV$, and both
supersymmetry breaking scenaria. The variability in the $\tilde\tau_{1,2}$,
$\tilde b_{1,2}$, and $\tilde t_{1,2}$ masses is because of the off-diagonal
elements of the corresponding mass matrices.}
\label{Figure3}
\end{figure}

The one-loop corrected lightest CP-even ($h$) and CP-odd ($A$) Higgs boson
masses are shown in Fig. \ref{Figure4} as functions of $m_{\tilde g}$ for
$m_t=150\GeV$. Following the methods of Ref. \cite{LNPWZh} we have determined
that the LEP lower bound on $m_h$ becomes $m_h\gsim60\GeV$, as the figure
shows. The largest value of $m_h$ depends on $m_t$; we find
\beq
m_h<\left\{	\begin{array}{ll}
		106,115,125\GeV;&\quad\vev{F_M}_{m_0=0}\\
		107,117,125\GeV;&\quad\vev{F_D}
			\end{array}
		\right.
\eeq
for $m_t=130,150,170\GeV$. It is interesting to note that the one-loop
corrected values of $m_h$ for $\tan\beta=2$ are quite dependent on the sign of
$\mu$. This phenomenon can be traced back to the $\tilde t_1-\tilde t_2$ mass
splitting which enhances the dominant $\tilde t$ one-loop corrections to $m_h$
\cite{ERZ}, an effect which is usually neglected in phenomenological analyses.
The $\tilde t_{1,2}$ masses for $\tan\beta=2$ and are drawn closer together
than the rest. The opposite effect occurs for $\mu<0$ and therefore the
one-loop correction is larger in this case. The sign-of-$\mu$ dependence
appears in the off-diagonal entries in the $\tilde t$ mass matrix  $\propto
m_t(A_t+\mu/\tan\beta)$, with $A_t<0$ in this case. Clearly only small
$\tan\beta$ matters, and $\mu<0$ enhances the splitting. The $A$-mass grows
fairly linearly with $m_{\tilde g}$ with a $\tan\beta$-dependent slope which
decreases for increasing $\tan\beta$, as shown in Fig. \ref{Figure4}. Note that
even though $m_A$ can be fairly light, we always get $m_A>m_h$, in agreement
with a general theorem to this effect in supergravity theories \cite{DNh}. This
result also implies that the channel $e^+e^-\to hA$ at LEPI is not
kinematically allowed in this model.

\begin{figure}[p]
\vspace{4.3in}
%\special{psfile=proc_erice4a.ps angle=90 hscale=60 vscale=55 hoffset=440}
\vspace{3.8in}
%\special{psfile=proc_erice4b.ps angle=90 hscale=60 vscale=55 hoffset=440}
\vspace{-0.3in}
\caption{\baselineskip=12pt
The one-loop corrected $h$ and $A$ Higgs masses versus the gluino
mass for both signs of $\mu$, $m_t=150\GeV$, and the two supersymmetry
breaking scenaria. Representative values of $\tan\beta$ are indicated.}
\label{Figure4}
\end{figure}

\subsection{Neutralino relic density}
The computation of the neutralino relic density (following the methods of
Refs. \cite{LNYdmI,KLNPYdm}) shows that $\Omega_\chi h^2_0\lsim0.25\,(0.90)$ in
the no-scale (dilaton) model. This implies that in these models the
cosmologically interesting values
$\Omega_\chi h^2_0\lsim1$ occur quite naturally. These results are in good
agreement with the observational upper bound on $\Omega_\chi h^2_0$ \cite{KT}.
Moreover, fits to the COBE data and the small and large scale structure of the
Universe suggest \cite{many} a mixture of $\approx70\%$ cold dark matter and
$\approx30\%$ hot dark matter together with $h_0\approx0.5$. The hot dark
matter component in the form of massive tau neutrinos has already been shown to
be compatible with the flipped $SU(5)$ model we consider here
\cite{chorus,ELNO}, whereas the cold dark matter component implies
$\Omega_\chi h^2_0\approx0.17$ which is reachable in these models.

\section{Flipped Phenomenology: Special Cases}
\label{PhenoSp}
\subsection{The strict no-scale case}
\label{strictNo-scale}
We now impose the additional constraint $B(M_U)=0$ to be added to
Eq.~(\ref{nsc}), and obtain the so-called strict no-scale case. Since $B(M_Z)$
is determined by the radiative electroweak symmetry breaking conditions, this
added constraint needs to beimposed in a rather indirect way. That is, for
given $m_{\tilde g}$ and $m_t$ values, we scan the possible values of
$\tan\beta$ looking for cases where $B(M_U)=0$. The most striking result is
that solutions exist {\em only} for $m_t\lsim135\GeV$ if $\mu>0$ and for
$m_t\gsim140\GeV$ if $\mu<0$. That is, the value of $m_t$ {\em determines} the
sign of $\mu$. Furthermore, for $\mu<0$ the value of $\tan\beta$ is determined
uniquely as a function of $m_t$ and $m_{\tilde g}$, whereas for $\mu>0$,
$\tan\beta$ can be double-valued for some $m_t$ range which includes
$m_t=130\GeV$ (but does not include $m_t=100\GeV$). In Fig. \ref{Figure5} (top
row) we plot the solutions found in this manner for the indicated $m_t$ values.

All the mass relationships deduced in the previous section apply here as well.
The $\tan\beta$-spread that some of them have will be much reduced though.
The most noticeable changes occur for the quantities which depend most
sensitively on $\tan\beta$. In Fig. \ref{Figure5} (bottom row) we plot the
one-loop corrected lightest Higgs boson mass versus $m_{\tilde g}$. The result
is that $m_h$ is basically determined by $m_t$; only a weak dependence on
$m_{\tilde g}$ exists. Moreover, for $m_t\lsim135\GeV\Leftrightarrow\mu>0$,
$m_h\lsim105\GeV$; whereas for $m_t\gsim140\GeV\Leftrightarrow\mu<0$,
$m_h\gsim100\GeV$. Therefore, in the strict no-scale case, once the top-quark
mass is measured, we will know the sign of $\mu$ and whether $m_h$ is above or
below $100\GeV$.

For $\mu>0$, we just showed that the strict no-scale constraint requires
$m_t\lsim135\GeV$. This implies that $\mu$ cannot grow as large as it did
previously in the general case. In fact, for $\mu>0$, $\mu_{max}\approx745\GeV$
before and $\mu_{max}\approx440\GeV$ now. This smaller value of $\mu_{max}$ has
the effect of cutting off the growth of the $\chi^0_{3,4},\chi^\pm_2$ masses
at $\approx\mu_{max}\approx440\GeV$ (c.f. $\approx750\GeV$) and of the heavy
Higgs masses at $\approx530\GeV$ (c.f. $\approx940\GeV$).

\begin{figure}[t]
\vspace{4.3in}
%\special{psfile=proc_erice5.ps angle=90 hscale=60 vscale=55 hoffset=440}
\vspace{-0.3in}
\caption{\baselineskip=12pt
The value of $\tan\beta$ versus $m_{\tilde g}$ in the strict no-scale case
(where $B(M_U)=0$) for the indicated values of $m_t$. Note that the sign of
$\mu$ is {\em determined} by $m_t$ and that $\tan\beta$ can be double-valued
for $\mu>0$. Also shown is the one-loop corrected lightest
Higgs boson mass. Note that if $\mu>0$ (for $m_t<135\GeV$) then $m_h<105\GeV$;
whereas if $\mu<0$ (for $m_t>140\GeV$) then $m_h>100\GeV$.}
\label{Figure5}
\end{figure}

\subsection{The special dilaton scenario case}
\label{specialDilaton}
In our analysis above, the radiative electroweak breaking conditions were used
to determine the magnitude of the Higgs mixing term $\mu$ at the electroweak
scale. This quantity is ensured to remain light as long as the supersymmetry
breaking parameters remain light. In a fundamental theory this parameter should
be calculable and its value used to determine the $Z$-boson mass. From this
point of view it is not clear that the natural value of $\mu$ should be light.
In specific models on can obtain such values by invoking non-renormalizable
interactions \cite{muproblem,Casasmu}. Another contribution to this quantity
is generically present in string supergravity models \cite{GM,Casasmu,KL}.
The general case with contributions from both sources has been effectively
dealt with in the previous section. If one assumes that only
supergravity-induced contributions to $\mu$ exist, then it can be shown that
the $B$-parameter at the unification scale is also determined \cite{KL},
\beq
B(M_U)=2m_0=\coeff{2}{\sqrt{3}}m_{1/2},\label{klII}
\eeq
which is to be added to the set of relations in Eq. (\ref{kl}). This new
constraint effectively determines $\tan\beta$ for given $m_t$ and $m_{\tilde
g}$ values and makes this restricted version of the model highly predictive.

{}From the outset we note that only solutions with $\mu<0$ exist. This is not
a completely obvious result, but it can be partially understood as follows.
In tree-level approximation, $m^2_A>0\Rightarrow\mu B<0$ at the electroweak
scale. Since $B(M_U)$ is required to be positive and not small, $B(M_Z)$ will
likely be positive also, thus forcing $\mu$ to be negative. A sufficiently
small value of $B(M_U)$ and/or one-loop corrections to $m^2_A$ could alter this
result, although in practice this does not happen. A numerical iterative
procedure allows us to determine the value of $\tan\beta$ which satisfies Eq.
(\ref{klII}), from the calculated value of $B(M_Z)$. We find that
\beq
\tan\beta\approx1.57-1.63,1.37-1.45,1.38-1.40
\quad{\rm for\ }m_t=130,150,155\GeV
\eeq
is required. Since $\tan\beta$ is so small ($m^{tree}_h\approx28-41\GeV$), a
significant one-loop correction to $m_h$ is required to increase it above
its experimental lower bound of $\approx60\GeV$ \cite{LNPWZh}. This requires
the largest possible top-quark masses (and a not-too-small squark mass).
However, perturbative unification imposes an upper bound on $m_t$ for a given
$\tan\beta$ \cite{DL}, which in this case implies \cite{aspects}
\beq
m_t\lsim155\GeV,
\eeq
which limits the magnitude of $m_h$
\beq
m_h\lsim74,87,91\GeV\qquad{\rm for}\qquad m_t=130,150,155\GeV.
\eeq
Lower values of $m_t$ are experimentally disfavored.

In Table~\ref{Table3} we give the range of sparticle and Higgs masses that
are allowed in this case. Clearly, continuing top-quark searches at the
Tevatron and Higgs searches at LEPI,II should probe this restricted scenario
completely.

\begin{table}
\hrule
\caption{
The range of allowed sparticle and Higgs masses in the restricted dilaton
scenario. The top-quark mass is restricted to be $m_t<155\GeV$. All masses in
GeV.}
\label{Table3}
\begin{center}
\begin{tabular}{|c|c|c|c|}\hline
$m_t$&$130$&$150$&$155$\\ \hline
$\tilde g$&$335-1000$&$260-1000$&$640-1000$\\
$\chi^0_1$&$38-140$&$24-140$&$90-140$\\
$\chi^0_2,\chi^\pm_1$&$75-270$&$50-270$&$170-270$\\
$\tan\beta$&$1.57-1.63$&$1.37-1.45$&$1.38-1.40$\\
$h$&$61-74$&$64-87$&$84-91$\\
$\tilde l$&$110-400$&$90-400$&$210-400$\\
$\tilde q$&$335-1000$&$260-1000$&$640-1000$\\
$A,H,H^+$&$>400$&$>400$&$>970$\\ \hline
 \end{tabular}
\end{center}
\hrule
\end{table}

%r \subsection{The light gluino scenario}

\section{The Minimal SU(5) Supergravity Model}
\label{minimal}
The minimal $SU(5)$ supergravity model \cite{Dickreview} needs to be specified
clearly in order to avoid the common misconception that it is simply the
so-called MSSM with the low-energy gauge couplings meeting at very high
energies. Two of its elements are particularly important: (i) it is a
supergravity model \cite{CAN} and as such the soft supersymmetry breaking
masses which allow unification are in principle calculable and are assumed to
be parametrized in terms of $m_{1/2},m_0,A$; and (ii) there exist
dimension-five proton decay operators \cite{WSY}, which are much larger than
the usual dimension-six operators, and require either a tuning of the
supersymmetry breaking parameters or a large Higgs triplet mass scale, to
obtain a sufficiently long proton lifetime
\cite{ENR,EMN,ANoldpd,MATS,ANpd,HMY,LNP,LNPZ}.

The $SU(5)$ symmetry is broken down to $SU(3)\times SU(2)\times U(1)$ via a
vev of the neutral component of the adjoint \r{24} of Higgs. The low-energy
pair of Higgs doublets are contained in the \r{5},\rb{5} Higgs representations.
Of the various proposals to split the proton-decay-mediating Higgs triplets
from the light Higgs doublets, perhaps the most appealing one is the so-called
``missing partner mechanism" \cite{MPM}, whereby a \r{75} of Higgs breaks the
gauge symmetry and the \r{5},\rb{5} pentaplets are coupled to \r{50},\rb{50}
representations (${\bf50}\cdot{\bf75}\cdot{\bf5}$,
$\ov{\bf50}\cdot{\bf75}\cdot\bar{\bf5}$). The doublets remain massless, while
the triplets acquire $\sim M_U$ masses. We should remark that this non-minimal
symmetry breaking mechanism is {\em not} the one that is usually considered
in studies of high-energy threshold effects in gauge coupling unification,
where one usually assumes that it is the \r{24} which effects the breaking.
\subsection{Gauge and Yukawa coupling unification}
This problem can be tackled at several levels of sophistication, which entail
an increasing number of additional assumptions. The most elementary approach
consists of running the one-loop supersymmetric gauge coupling RGEs starting
with the precisely measured values of $\alpha_e,\alpha_3,\sin^2\theta_w$ at
the scale $M_Z$ and discovering that the three gauge couplings meet at the
scale $M_U\sim10^{16}\GeV$ \cite{CostaAmaldi}. More interesting from the
theoretical standpoint is to assume that unification must occur, as is the case
in the minimal $SU(5)$ supergravity model, and use this constraint to predict
the low-energy value of $\sin^2\theta_w$ in terms of $\alpha_e$ and $\alpha_3$.
The next level of sophistication consists of increasing the accuracy of the
RGEs to two-loop level and parametrize the supersymmetric threshold by a single
mass parameter between $\sim M_Z$ and $\sim{\rm few}\TeV$
\cite{EKNI,LL,Arason,AnselmoI}. More realistically, one specifies the whole
light supersymmetric spectrum in detail
\cite{EKNII,EKNIII,EKNIV,RR,HMY,AnselmoIII,AnselmoIV,AnselmoVI,LP,CPW},
as well as some subtle effects such as the evolution of the gaugino masses
(EGM) effect \cite{AnselmoIII,EKNIV}, and the effect of the Yukawa couplings on
the two-loop gauge coupling RGEs \cite{BBO}. A final step of sophistication
attempts to model the transition from the $SU(3)\times SU(2)\times U(1)$ theory
onto the $SU(5)$ theory by means of high-energy threshold effects which depend
on the masses of the  various GUT fields
\cite{EKNIII,EKNIV,BH,HMY,AnselmoVI,LP} as well as on
coefficients of possible non-renormalizable operators \cite{LP,HS}. This last
step does away completely with the concept of a single ``unification mass". In
fact, until this last step is actually accounted for somehow, one is not
dealing with a true unified theory since otherwise the gauge couplings would
diverge again past the unification scale, \ie, ``physics is not euclidean
geometry".

It is interesting to note that the original hope that precise knowledge of
the low energy gauge couplings would constrain the scale of the low-energy
supersymmetric particles, has not bear fruit \cite{BH,AnselmoIII,EKNIV}, mainly
because of the largely unknown GUT threshold effects. More precisely, the
supersymmetric particle masses can lie anywhere up to $\sim{\rm few}\TeV$
provided the parameters of the GUT theory are adjusted accordingly.

Another consequence of the $SU(5)$ symmetry is the relation $\lambda_b(M_U)=
\lambda_\tau(M_U)$ which when renormalized down to low energies gives a ratio
$m_b/m_\tau$ in fairly good agreement with experiment \cite{BEGN}.  This
problem can also be tackled with improving degree of sophistication
\cite{EKNI,Arason,EKNIII,GHS,YU,DHR,BBO,Naculich,LPII} and even postulating
some high-energy threshold effects \cite{BBO,LPII}.
%r However, the need to make contact with the poorly understood quantity
%r $m_b(m_b)$, makes the recent intense efforts to push this analysis to the
%r limit, perhaps somewhat of an overkill.
In practice, the
$\lambda_b(M_U)=\lambda_\tau(M_U)$ constraint entails a relationship between
$m_t$ and $\tan\beta$, \ie, $\tan\beta=\tan\beta(m_t,m_b,\alpha_3)$, as
follows: (i) the values of $m_b$ and $m_\tau$, together with $\tan\beta$
determine the low-energy values of $\lambda_b$ and $\lambda_\tau$; (ii) the
input value of $m_t$ determines the low-energy value of $\lambda_t$; (iii)
running these three Yukawa couplings up to the unification scale one discovers
the above relation between $\tan\beta$ and $m_t$ if the Yukawa unification
constraint is satisfied. In actuality, the dependence on $m_b$ and $\alpha_3$
is quite important. We note that for arbitrary choices of $m_t$ and
$\tan\beta$, one obtains values of $m_b$ typically close to or above $5\GeV$,
whereas popular belief would like to see values below $4.5\GeV$. Strict
adherence to this prediction for $m_b$ requires that one be in a rather
constrained region of the $(m_t,\tan\beta)$ plane, where $\tan\beta\sim1$
or $\gsim40$ \cite{EKNIV,YU,BBO,LPII}, or that $m_t$ be large (above
$180\GeV$). We do not impose this stringent constraint on the parameter space,
hoping that further contributions to the quark masses (as required in $SU(5)$
GUTs to fit the lighter generations also \cite{EG}) will relax it somehow.

As noted above, the parameter space of this model can be described in terms
of five parameters: $m_t,\tan\beta,m_{1/2},m_0,A$. In Ref. \cite{LNP} we
performed an exploration of the following hypercube of the parameter space:
$\mu>0,\mu<0$, $\tan\beta=2-10\,(2)$, $m_t=100-160\,(5)$, $\xi_0=0-10\,(1)$,
$\xi_A=-\xi_0,0,+\xi_0$, and $m_{1/2}=50-300\,(6)$, where the numbers in
parenthesis represent the size of the step taken in that particular direction.
(Points outside these ranges have little (a posteriori) likelihood of being
acceptable.) Of these $92,235\times2=184,470$ points, $\approx25\%$ passed
all the standard constraints, \ie, radiative electroweak symmetry breaking
and all low-energy phenomenology as described in Ref. \cite{aspects}. The
most important constraint on this parameter space is proton decay, as discussed
below. First we discuss some aspects of the gauge coupling unification
calculation.

As a first step we used one-loop gauge coupling RGEs and a common
supersymmetric threshold at $M_Z$, to determine $M_U,\alpha_U$, and
$\sin^2\theta_w$, once $\alpha_3(M_Z)=0.113,0.120$ and
$\alpha^{-1}_e(M_Z)=127.9$ were given. In Ref. \cite{LNPZ} we refined our
study including several important features: (i) recalculation of $M_U$ using
two-loop gauge coupling RGEs including light supersymmetric thresholds, (ii)
exploration of values of $\alpha_3$ throughout its $\pm1\sigma$ allowed range,
and (iii) exploration of low values of $\tan\beta\,(<2)$ (which maximize the
proton lifetime). We used the analytical approximations to the solution of the
two-loop gauge coupling RGEs in Ref. \cite{AnselmoIII} to obtain
$M_U,\alpha_U,\sin^2\theta_w$. The supersymmetric threshold was treated in
great detail \cite{AnselmoIII} with all the sparticle masses obtained from our
procedure \cite{LNP}. Since, the sparticle masses vary as one explores the
parameter space, one obtains {\it ranges} for the calculated values. In
Table~\ref{Table5} we show the one-loop value for $M_U$ ($M_U^{(0)}$), the
two-loop plus supersymmetric threshold corrected unification mass range
($M_U^{(1)}$) [as expected \cite{AnselmoIII} $M_U$ is reduced by both effects],
the ratio of the two, and the calculated range of $\sin^2\theta_w$.\footnote{
We should note that these ranges are obtained after all constraints discussed
below have been satisfied, the proton decay being the most important one.}
Note that for $\alpha_3=0.118$ (and lower), $\sin^2\theta_w$ is outside the
experimental $\pm1\sigma$ range ($\sin^2\theta_w=0.2324\pm0.0006$ \cite{LP}),
whereas $\alpha_3=0.126$ gives quite acceptable values.

\begin{table}[t]
\hrule
\caption{
The value of the one-loop unification mass $M_U^{(0)}$, the two-loop and
supersymmetric threshold corrected unification mass range $M_U^{(1)}$, the
ratio of the two, and the range of the calculated $\sin^2\theta_w$, for the
indicated values of $\alpha_3$ (the superscript $+\,(-)$ denotes $\mu>0\,(<0)$)
and $\alpha^{-1}_e=127.9$. The $\sin^2\theta_w$ values should be compared with
the current experimental $\pm1\sigma$ range $\sin^2\theta_w=0.2324\pm0.0006$
[95]. Lower values of $\alpha_3$ drive $\sin^2\theta_w$ to values even higher
than for $\alpha_3=0.118$. All masses in units of $10^{16}\GeV$.}
\label{Table5}
\begin{center}
\begin{tabular}{|c||c|c||c|c|}\hline
&$\alpha_3=0.126^+$&$\alpha_3=0.126^-$&$\alpha_3=0.118^+$&$\alpha_3=0.118^-$
\\ \hline
$M^{(0)}_U$&$3.33$&$3.33$&$2.12$&$2.12$\\
$M^{(1)}_U$&$1.60-2.13$&$1.60-2.05$&$1.02-1.35$&$1.02-1.30$\\
$M_U^{(1)}/M_U^{(0)}$&$0.48-0.64$&$0.48-0.61$&$0.48-0.64$&$0.48-0.61$\\
$\sin^2\theta_w$&$0.2315-0.2332$&$0.2313-0.2326$&$0.2335-0.2351$&
$0.2332-0.2345$\\ \hline
\end{tabular}
\end{center}
\hrule
\end{table}

We do not specify the details of the GUT thresholds and in practice take two
of the GUT mass parameters (the masses of the $X,Y$ gauge bosons $M_V$, and
the mass of the adjoint Higgs multiplet $M_\Sigma$) to be degenerate with
$M_U$. Since below we allow $M_H<3M_U$, Table~\ref{Table5} indicates that in
our calculations $M_H<6.4\times10^{16}\GeV$. In Ref. \cite{HMY} it is argued
that a more proper upper bound is $M_H<2M_V$, but $M_V$ cannot be calculated
directly, only $(M^2_V M_\Sigma)^{1/3}<3.3\times10^{16}\GeV$ is known from
low-energy data \cite{HMY}. If we take $M_\Sigma=M_V$, this would give
$M_H<2M_V<6.6\times10^{16}\GeV$, which agrees with our present requirement.
Below we comment on the case $M_\Sigma<M_V$.
\subsection{Proton decay}
\label{su5pdecay}
In the minimal $SU(5)$ supergravity model only the dimension-five--mediated
proton decay operators are constraining. In calculating the proton lifetime we
consider the typically dominant decay modes $p\to \bar\nu_{\mu,\tau} K^+$ and
neglect all other possible modes. Schematically the lifetime is given
by\footnote{Throughout our calculations we have used the explicit proton
decay formulas in Ref. \cite{ANoldpd}.}
\beq
\tau_p\equiv\tau(p\to\bar\nu_{\mu,\tau}K^+)
\sim\left| M_H \sin2\beta {1\over f}{1\over 1+y^{tK}}\right|^2.
\eeq
Here $M_H$ is the mass of the exchanged GUT Higgs triplet which on perturbative
grounds is assumed to be bounded above by $M_H<3M_U$ \footnote{This relation
assumes implicitly that all the components of the \r{24} superfield are nearly
degenerate in mass \cite{HMY}.} \cite{EMN,ANpd,HMY};
$\sin2\beta=2\tan\beta/(1+\tan^2\beta)$, thus $\tau_p$ `likes' small
$\tan\beta$ (we find that only $\tan\beta\lsim6$ is allowed); $y^{tK}$
represents the calculable ratio of the third- to the second-generation
contributions to the dressing one-loop diagrams. An unkown phase appears in
this ratio (which has generally $|y^{tK}|\ll1$) and we always consider the
weakest possible case of  destructive interference. Finally $f$ represents the
sparticle-mass--dependent dressing one-loop function which decreases
asymptotically with large sparticle masses.

In Fig.~\ref{Figure11} (top row) we show a scatter plot of $(\tau_p,m_{\tilde
g})$. The various `branches' correspond to fixed values of $\xi_0$. Note that
for $\xi_0<3$, $\tau_p<\tau^{exp}_p=1\times10^{32}\y$ (at 90\% C.L.
\cite{PDG}). Also, for a given value of $\xi_0$, there is a corresponding
allowed interval in $m_{\tilde g}$. The lower end of this interval is
determined by the fact that $\tau_p\propto1/f^2$, and $f\approx
m_{\chi^+_1}/m^2_{\tilde q}\propto 1/m_{\tilde g}(c+\xi^2_0)$, in the
proton-decay--favored limit of $\mu\gg M_W$; thus
$m_{\tilde g}(c+\xi^2_0)>{\rm constant}$. The upper end of the interval follows
from our requirement $m_{\tilde q}(\propto m_{\tilde
g}\sqrt{c+\xi^2_0})<1\TeV$. Statistically speaking, the proton decay cut is
quite severe, allowing only about $\sim1/10$ of the points which passed all the
standard constraints, independently of the sign of $\mu$.

Note that if we take $M_H=M_U$ (instead of $M_H=3M_U$), then
$\tau_p\to{1\over9}\tau_p$ and all points in Fig.~\ref{Figure11} would become
excluded. To obtain a rigorous lower bound on $M_H$, we would need to explore
the lowest possible allowed values of $\tan\beta$ (in Fig.~\ref{Figure11},
$\tan\beta\ge2$). Roughly, since the dominant $\tan\beta$ dependence of
$\tau_p$ is through the explicit $\sin2\beta$ factor, the upper bound
$\tau_p\lsim8\times10^{32}\y$ for $\tan\beta=2$, would become
$\tau_p\lsim1\times10^{33}\y$ for $\tan\beta=1$. Therefore, the current
experimental lower bound on $\tau_p$ would imply $M_H\gsim M_U$. Note also that
SuperKamiokande ($\tau^{exp}_p\approx2\times10^{33}\y$) should be able to probe
the whole allowed range of $\tau_p$ values.

The actual value of $\alpha_3(M_Z)$ used in the calculations ($\alpha_3=0.120$
in Fig.~\ref{Figure11}) has a non-negligible effect of some of the final
results, mostly due to its effect on the value of $M_H=3M_U$: larger values of
$\alpha_3$ increase $M_U$ and therefore $\tau_p$, and thus open up the
parameter space, and viceversa. For example, for $\alpha_3=0.113\,(0.120)$ we
get $m_{\tilde g}\lsim550\,(800)\GeV$, $\xi_0\ge5\,(3)$, and
$\tau_p\lsim4\,(8)\times10^{32}\y$.

\begin{figure}[p]
\vspace{4.9in}
%\special{psfile=proc_summer11.ps angle=90 hscale=65 vscale=63 hoffset=460}
\vspace{-0.3in}
\caption{\baselineskip=12pt
Scatter plot of the proton lifetime
$\tau_p\equiv\tau(p\to\bar\nu_{\mu,\tau}K^+)$ versus the gluino mass for the
hypercube of the parameter space explored. The unification mass is calculated
in one-loop approximation assuming a common supersymmetric threshold at $M_Z$,
and $M_H=3M_U$ is assumed. The current experimental lower bound  is
$\tau^{exp}_p=1\times10^{32}\y$. The various `branches' correspond to fixed
values of $\xi_0$ as indicated (the labelling applies to all four windows). The
bottom row includes the cosmological constraint. The upper bound on $m_{\tilde
g}$ follows from the requirement $m_{\tilde q}<1\TeV$.}
\label{Figure11}
\end{figure}

\begin{figure}[t]
\vspace{4.3in}
%\special{psfile=proc_summer12.ps angle=90 hscale=60 vscale=55 hoffset=440}
\vspace{-1.5in}
\caption{\baselineskip=12pt
The calculated values of the proton lifetime into $p\to\bar\nu K^+$
versus the lightest chargino (or second-to-lightest neutralino) mass for both
signs of $\mu$, using the more accurate value of the unification mass (which
includes two-loop and low-energy supersymmetric threshold effects). Note that
we have taken $\alpha_3+1\sigma$ in order to maximize $\tau_p$. Note also that
future proton decay experiments should be sensitive up to
$\tau_p\approx20\times10^{32}\y$.}
\label{Figure12}
\end{figure}

The refinement on the calculation of the unification mass described above, to
include two-loop effects and light supersymmetric thresholds, has a significant
effect on the calculated value of the proton lifetime \cite{LNPZ}, since we
take $M_H=3M_U$. With the more accurate value of $M_U$ we  simply rescale our
previously calculated
$\tau_p$ values which satisfied $\tau^{(0)}_p>\tau^{exp}_p$, and find that
$\tau^{(1)}_p=\tau^{(0)}_p[M_U^{(1)}/M_U^{(0)}]^2>\tau^{exp}_p$ for only
$\lsim25\%$ of the previously allowed points. The value of $\alpha_3$ has a
significant influence on the results since (see Table~\ref{Table5}) larger
(smaller) values of $\alpha_3$ increase (decrease) $M_U$, although the effect
is more pronounced for low values of $\alpha_3$. To quote the most conservative
values of the observables, in what follows we take $\alpha_3$ at its $+1\sigma$
value ($\alpha_3=0.126$). As discussed above, this choice of $\alpha_3$ also
gives $\sin^2\theta_w$ values consistent with the $\pm1\sigma$ experimental
range. Finally, in the search  of the parameter space above, we considered only
$\tan\beta=2,4,6,8,10$ and found that $\tan\beta\lsim6$ was required. Our
present analysis indicates that this upper bound is reduced down to
$\tan\beta\lsim3.5$. Here we consider also $\tan\beta=1.5,1.75$ since low
$\tan\beta$ maximizes $\tau_p\propto \sin^22\beta$. These add new allowed
points (\ie, $\tau^{(0)}_p>\tau^{exp}_p$) to our previous set, although most of
them ($\gsim75\%$) do not survive the stricter proton decay constraint
($\tau^{(1)}_p>\tau^{exp}_p$) imposed here.

In Fig.~\ref{Figure12} we show the re-scaled values of $\tau_p$ versus the
lightest chargino mass $m_{\chi^\pm_1}$. All points satisfy $\xi_0\equiv
m_0/m_{1/2}\gsim6$ and $m_{\chi^\pm_1}\lsim150\GeV$, which are to be contrasted
with $\xi_0\gsim3$ and $m_{\chi^\pm_1}\lsim225\GeV$ derived using
the {\em weaker} proton decay constraint \cite{LNP}. The upper bound
on $m_{\chi^\pm_1}$ derives from its near proportionality to $m_{\tilde g}$,
$m_{\chi^\pm_1}\approx0.3m_{\tilde g}$ \cite{ANpd,LNP}, and the result
$m_{\tilde g}\lsim500\GeV$. The latter follows from the proton decay constraint
$\xi_0\gsim6$ and the naturalness requirement $m_{\tilde q}\approx\sqrt{m^2_0+
6m^2_{1/2}}\approx{1\over3}m_{\tilde g}\sqrt{6+\xi^2_0}<1\TeV$. Within our
naturalness and $H_3$ mass assumptions, we then obtain \footnote{Note that in
general, $\tau_p\propto M^2_{H_3}[m^2_{\tilde q}/
m_{\chi^\pm_1}]^2\propto M^2_{H_3}[m_{\tilde g}(6+\xi^2_0)]^2$ and thus
$\tau_p$ can be made as large as desired by increasing sufficiently either
the supersymmetric spectrum or $M_H$.}
\beq
\tau_p<3.1\,(3.4)\times10^{32}\y\quad{\rm for}\quad \mu>0\,(\mu<0).\label{taup}
\eeq
The $p\to\bar\nu K^+$ mode should then be readily observable at SuperKamiokande
and Gran Sasso since these experiments should be sensitive up to
$\tau_p\approx2\times10^{33}\y$. Note that if $M_H$ is relaxed up to its
largest possible value consistent with low-energy physics,
$M_H=2.3\times10^{17}\GeV$ \cite{HMY}, then in Eq.~(\ref{taup}) $\tau_p\to
\tau_p<4.0\,(4.8)\times10^{33}\y$, and only part of the parameter space of
the model would be experimentally accessible. However, to make this choice
of $M_H$ consistent with high-energy physics (\ie, $M_H<2M_V$) one must
have $M_V/M_\Sigma>42$.

\subsection{Neutralino relic density}
\label{su5dm}
The study of the relic density of neutralinos requires the knowledge of the
total annihilation amplitude $\chi\chi\to all$. The latter depends on the model
parameters to determine all masses and couplings. Previously
\cite{LNYdmI,KLNPYdm} we have advocated the study of this problem in the
context of supergravity models with radiative electroweak symmetry breaking,
since then only a few parameters (five or less) are needed to specify the model
completely. In particular, one can explore the whole parameter space and draw
conclusions about a complete class of models. The ensuing relationships among
the various masses and couplings have been found to yield results which depart
from the conventional minimal supersymmetric standard model (MSSM) lore, where
no such relations exist. In the minimal $SU(5)$ supergravity model we have just
shown that its five-dimensional parameter space is strongly constrained by the
proton lifetime. It was first noticed in Ref.~\cite{troubles} that the
neutralino relic density for the proton-decay allowed points in parameter space
is large, \ie, $\Omega_\chi h^2_0\gg1$, and therefore in conflict with current
cosmological expectations: requiring that the Universe be older than the oldest
known stars implies $\Omega_0 h^2_0\le1$ \cite{KT}.

In Refs.~\cite{troubles,LNP,LNPZ} the neutralino relic density has been
computed following the methods of Refs.~\cite{LNYdmI,KLNPYdm}. In
Fig.~\ref{Figure11} (bottom row) we show the effect of the cosmological
constraint on the parameter space allowed by proton decay. Only $\sim1/6$ of
the points satisfy $\Omega_\chi h^2_0\le1$. This result is not unexpected since
proton decay is  suppressed by heavy sparticle masses, whereas $\Omega_\chi
h^2_0$ is enhanced. Therefore, a delicate balance needs to be attained to
satisfy both constraints simultaneously. Note that the subset of cosmologically
allowed points does not change the range of possible $\tau_p$ values, although
it depletes the constant-$\xi_0$ `branches'.

The effect of the cosmological constraint is perhaps more manifest when one
considers the correlation between $m_h$ and $m_{\chi^\pm_1}$ after imposing
the (\eg, {\em weaker}) proton decay constraint, but with and without imposing
the cosmological constraint. This contrast is shown in Fig.~\ref{Figure14}.

\begin{figure}[t]
\vspace{5in}
%\special{psfile=proc_summer14.ps angle=90 hscale=63 vscale=63 hoffset=450}
\vspace{-0.3in}
\caption{\baselineskip=12pt
The allowed region in parameter space which satisfies the {\em weaker} proton
decay constraint, before and after the imposition of the cosmological
constraint. Note that when the cosmological constraint is imposed (bottom
row), an interesting correlation between the two particle masses arises.}
\label{Figure14}
\end{figure}

The main conclusion is that the relic density can be small only near the
$h$- and $Z$-pole resonances, \ie, for $m_\chi\approx{1\over2}m_{h,Z}$
\cite{LNP,ANcosm}, since in this case the annihilation cross section is
enhanced. It is important to note that in this type of calculations the thermal
average of the annihilation cross section is usually computed using an
expansion around threshold (\ie, $\sqrt{s}=2m_\chi$) \cite{SWO}. In Ref.
\cite{GG} however, it has been pointed out that the resulting thermal
average can be quite inaccurate near poles and thresholds of the annihilation
cross section, which is precisely the case for the points of interest in the
minimal $SU(5)$ model. In Ref.~\cite{ANcosm,poles} the relic density
calculation has been redone following the more accurate methods of
Ref.~\cite{GG}. The result is that the poles are broader and shallower, and
thus the cosmological constraint is weakened with respect to the standard
(using the expansion) procedure of performing the thermal average.
However, qualitatively the cosmologically allowed region of parameter space
is not changed. This result is shown in Fig.~\ref{Figure13}, where the
points in parameter space allowed by the stricter proton decay constraint and
cosmology are shown in the $(m_{\chi^\pm_1},m_h)$ plane.

\begin{figure}[p]
\vspace{4.0in}
%\special{psfile=proc_summer13a.ps angle=90 hscale=60 vscale=60 hoffset=450}
\vspace{2.5in}
%\special{psfile=proc_summer13b.ps angle=90 hscale=62 vscale=50 hoffset=450}
\vspace{-1in}
\caption{\baselineskip=12pt
The points in parameter space of the minimal $SU(5)$ supergravity model which
satisfy the stricter proton decay constraint and the cosmological constraint
with the relic density computed in approximate and accurate way. Note the
little qualitative difference between the two sets of plots.}
\label{Figure13}
\end{figure}

\subsection{Mass ranges and relations}
Since the proton decay constraint generally requires $|\mu|\gg M_W$ (and to a
somewhat lesser extent also $|\mu|\gg M_2$), the lightest chargino will have
mass $m_{\chi^+_1}\approx M_2\approx0.3m_{\tilde g}$, whereas the two lightest
neutralinos will have masses $m_\chi\approx M_1\approx{1\over2}M_2$ and
$m_{\chi^0_2}\approx M_2$ \cite{ANpd,ANc}. Thus, within some approximation we
expect the relation in Eq.~(\ref{neuchar}) to be satisfied in this model also.
Inclusion of the cosmological constraint does not affect significantly the
range of sparticle masses in Sec.~\ref{su5pdecay}. The value of $\alpha_3$ does
not affect these mass relations either, although the particle mass ranges do
change
\beq
m_\chi<85\,(115)\GeV,\qquad m_{\chi^0_2,\chi^+_1}<165\,(225)\GeV,
\qquad{\rm for}\ \alpha_3=0.113\,(0.120).
\eeq
The reason is simple: higher values of $\alpha_3$ increase $M_U$ and therefore
$M_H\,(=3M_U)$, which in turn weakens the proton decay constraint.
We also find that the one-loop corrected lightest Higgs boson mass ($m_h$) is
bounded above by
\beq
m_h\lsim110\,(100)\GeV,
\eeq
independently of the sign of $\mu$, the value of $\alpha_3$, or the
cosmological constraint; the stronger bound holds when the stricter proton
decay constraint is enforced. In Fig.~\ref{Figure13} we have shown $m_h$ versus
$m_{\chi^+_1}$ for $\tan\beta=1.5,1.75,2$; for the maximum allowed $\tan\beta$
value ($\approx3.5$), $m_h\lsim100\GeV$. Note that for $\mu>0$,
$m_h\approx50\GeV$, and $m_{\chi^\pm_1}\gsim100\GeV$, there is a sparsely
populated area with highly fine-tuned points in parameter space
($m_t\approx100\GeV$, $\tan\beta\approx1.5$, $\xi_A\equiv A/m_{1/2}
\approx\xi_0\approx6$). This figure shows an experimentally interesting
correlation when the cosmological constraints are imposed,
\beq
m_h\gsim72\GeV \Rightarrow m_{\chi^+_1}\lsim100\GeV.\label{mV}
\eeq
The bands of points towards low values of $m_h$ represent the discrete choices
of $\tan\beta=1.5,1.75$. The voids between these bands are to be understood as
filled by points with $1.5\lsim\tan\beta\lsim1.75$. For
$m_{\chi^\pm_1}>106\,(92)\GeV$ (for $\mu>0\,(\mu<0)$), we
obtain $m_h\lsim50\,(56)\GeV$ and Higgs detection at LEP should be immediate.
The correlations among the lightest chargino and neutralino masses imply
analogous results for $(m_h,m_{\chi^0_2})$ and $(m_h,m_\chi)$,
\beq
m_h\gsim80\GeV \Rightarrow m_{\chi^0_2}\lsim90\,(110)\GeV,
\quad m_\chi\lsim48\,(60)\GeV,\label{mVI}
\eeq
for $\alpha_3=0.113\,(0.120)$.  These correlations can be understood in the
following way: since we find that $m_A\gg M_Z$, then
$m_h\approx|\cos2\beta|M_Z+({\rm rad.\ corr.})$. In the
situation we consider here, we have determined that all of the allowed points
for $m_{\tilde g}>400\;\GeV$ correspond to $\tan\beta=2$. This implies that
the tree-level contribution to $m_h$ is $\approx55\GeV$. We also find that the
cosmology cut restricts  $m_t<130(140)\;\GeV$ for $\mu>0(\mu<0)$ in this range
of $m_{\tilde g}$. Therefore, the radiative correction contribution to $m^2_h$
($\propto m^4_t$) will be modest in this range of $m_{\tilde g}$. This explains
the depletion of points for $m_h\gsim 80\;\GeV$ in Fig.~\ref{Figure13} and
leads to the mass relationships in Eqs.~(\ref{mV},\ref{mVI}).

In this model the only light particles are the lightest Higgs boson
($m_h\lsim100\GeV$), the two lightest neutralinos ($m_{\chi^0_1}\approx
{1\over2}m_{\chi^0_2}\lsim75\GeV$), and the lightest chargino ($m_{\chi^\pm_1}
\approx m_{\chi^0_2}\lsim150\GeV$). The gluino and the lightest stop can be
light ($m_{\tilde g}\approx 160-460\GeV$, $m_{\tilde t_1}\approx170-825\GeV$),
but for most of the parameter space are not within the reach of Fermilab.

In Ref. \cite{LNPWZh} it has been shown that the actual LEP lower bound on the
ligthest Higgs boson mass is improved in the class of supergravity models
with radiative electroweak symmetry breaking which we consider here, one gets
$m_h\gsim60\GeV$. In Sec.~\ref{LEPI} below we discuss the details of this
procedure. For now it suffices to note that the improved bound $m_h\gsim60\GeV$
mostly restricts low values of $\tan\beta$ and therefore the minimal $SU(5)$
supergravity model where $\tan\beta\lsim3.5$ \cite{LNPZ}. Above we obtained
upper bounds on the light particle masses in this model ($\tilde
g,h,\chi^0_{1,2},\chi^\pm_1$) for $m_h>43\GeV$. In particular, it was found
that $m_{\chi^\pm_1}\gsim100\GeV$ was only possible for $m_h\lsim50\GeV$. The
improved bound on $m_h$ immediately implies the following considerably stronger
upper bounds
\begin{eqnarray}
m_{\chi^0_1}&\lsim&52(50)\GeV, \\
m_{\chi^0_2}&\lsim&103(94)\GeV,\\
m_{\chi^\pm_1}&\lsim&104(92)\GeV,\\
m_{\tilde g}&\lsim&320\,(405)\GeV,
\end{eqnarray}
for $\mu>0$ ($\mu<0$).

A related consequence is that the mass relation $m_{\chi^0_2}>m_{\chi^0_1}+m_h$
is not satisfied for any of the remaining points in parameter space
and therefore the $\chi^0_2\to\chi^0_1 h$ decay mode is not kinematically
allowed. Points where such mode was previously allowed  led to a vanishing
trilepton signal in the reaction $p\bar p\to\chi^\pm_1\chi^0_2$ at Fermilab
(thus the name `spoiler mode') \cite{LNWZ}. The improved situation now implies
at least one event per $100\ipb$ for all remaining points in parameter space
(see Sec.~\ref{tevatron}).

\section{Prospects for Direct Experimental Detection}
\label{Direct}
The sparticle and Higgs spectrum discussed in
Secs.~\ref{PhenoGen},\ref{PhenoSp},\ref{minimal} can be directly explored
partially at present and near future collider facilities, as we now discuss for
each model considered above.

\subsection{Tevatron}
\label{tevatron}
\begin{description}
\item (a) The search and eventual discovery of the top quark will narrow down
the parameter space of these models considerably. Moreover, in the two special
flipped $SU(5)$ cases discussed in Sec.~\ref{PhenoSp} this measurement
will be very important: (i) in the strict no-scale case
(Sec.~\ref{strictNo-scale}) it will determine the sign of $\mu$ ($\mu>0$ if
$m_t\lsim135\GeV$; $\mu<0$ if $m_t\gsim140\GeV$) and whether the Higgs mass is
above or below $\approx100\GeV$, and (ii) it may rule out the restricted
dilaton scenario (Sec.~\ref{specialDilaton}) if $m_t>150\GeV$.
\item (b) The trilepton signal in $p\bar p\to \chi^0_2\chi^\pm_1X$, where
$\chi^0_2$ and $\chi^\pm_1$ both decay leptonically, is a clean test of
supersymmetry \cite{trileptons} and in particular of this class of models
\cite{LNWZ}.  The trilepton rates in the no-scale flipped $SU(5)$ and in the
minimal $SU(5)$ models have been given in Ref. \cite{LNWZ}; in Fig.
\ref{Figure6} we show these (in the no-scale case $m_t=130\GeV$ has been
chosen). One can show that with ${\cal L}=100\ipb$ of integrated luminosity
basically all of the parameter space of the minimal $SU(5)$ model should be
explorable. Also, chargino masses as high as $\approx175\GeV$ in the no-scale
model could be explored, although some regions of parameter space for lighter
chargino masses would remain unexplored. We expect that somewhat weaker results
will hold for the dilaton model, since the sparticle masses are heavier in that
model, especially the sleptons which enhance the leptonic branching ratios when
they are light enough \cite{LNWZ}.
\item (c) The relation $m_{\tilde q}\approx m_{\tilde g}$ for the $\tilde
u_{L,R},\tilde d_{L,R}$ squark masses in the flipped $SU(5)$ models
should allow to probe the low end of the squark and gluino allowed mass ranges,
although the outlook is more promising for the dilaton model since the allowed
range starts off at lower values of $m_{\tilde g,\tilde q}$ (see Eq.
(\ref{gmin})). An important point distinguishing the two models is that the
average squark mass is slightly below (above) the gluino mass in the no-scale
(dilaton) model, which should have an important bearing on the experimental
signatures and rates \cite{sgdetection}. In the dilaton case the $\tilde t_1$
mass can be below $100\GeV$ for sufficiently low $m_t$, and thus may be
detectable. As the lower bound on $m_t$ rises, this signal becomes less
accessible. The actual reach of the Tevatron for the above processes depends on
its ultimate integrated luminosity. The squark masses in the minimal $SU(5)$
model ($m_{\tilde q}\gsim500\GeV$) are beyond the reach of the Tevatron.
\end{description}

\begin{figure}[p]
\vspace{4.5in}
%\special{psfile=proc_summer6a.ps angle=90 hscale=60 vscale=60 hoffset=440}
\vspace{3in}
%\special{psfile=proc_erice6.ps angle=90 hscale=60 vscale=60 hoffset=440}
\vspace{-2.2in}
\caption{\baselineskip=12pt
The number of trilepton events at the Tevatron per $100\ipb$ in the minimal
$SU(5)$ model and the no-scale flipped $SU(5)$ model (for $m_t=130\GeV$). Note
that with $200\ipb$ and 60\% detection efficiency it should be possible to
probe basically all of the parameter space of the minimal $SU(5)$ model, and
probe chargino masses as high as $175\GeV$ in the no-scale model.}
\label{Figure6}
\end{figure}

\subsection{LEPI}
\label{LEPI}
The current LEPI lower bound on the Standard Model (SM) Higgs boson mass
($m_H>61.6\GeV$ \cite{Hilgart}) is obtained by studying the
process $e^+e^-\to Z^*H$ with subsequent Higgs decay into two jets. The MSSM
analog of this production process leads to a cross section differing just by a
factor of $\sin^2(\alpha-\beta)$, where $\alpha$ is the SUSY Higgs mixing angle
and $\tan\beta=v_2/v_1$ is the ratio of the Higgs vacuum expectation values
\cite{HHG}. The published LEPI lower bound on the lightest SUSY Higgs boson
mass ($m_h>43\GeV$) is the result of allowing $\sin^2(\alpha-\beta)$ to vary
throughout the MSSM parameter space and by considering the $e^+e^-\to Z^*h,hA$
cross sections. It is therefore possible that in specific models (which embed
the MSSM), where $\sin^2(\alpha-\beta)$ is naturally restricted to be near
unity, the lower bound on $m_h$ could rise, and even reach the SM lower bound
if ${\rm BR}(h\to2\,{\rm jets})$ is SM-like as well. This has been shown to
be the case for the supergravity models we discuss here, and more generally
for supergravity models which enforce radiative electroweak symmetry
breaking \cite{LNPWZh}.

Non-observation of a SM Higgs signal puts the following upper bound in the
number of expected 2-jet events.
\beq
\#{\rm events}_{\rm\,SM}=\sigma(e^+e^-\to Z^*H)_{\rm SM}\times
{\rm BR}(H\to2\,{\rm jets})_{\rm SM}\times\int{\cal L}dt<3.\label{LI}
\eeq
The SM value for ${\rm BR}(H\to2\,{\rm jets})_{\rm SM}\approx
{\rm BR}(H\to b\bar b+c\bar c+gg)_{\rm SM}\approx0.92$ \cite{HHG} corresponds
to an upper bound on $\sigma(e^+e^-\to Z^*H)_{\rm SM}$. Since this is a
monotonically decreasing function of $m_H$, a lower bound on $m_H$ follows,
\ie, $m_H>61.6\GeV$ as noted above. We denote by $\sigma_{\rm SM}(61.6)$ the
corresponding value for $\sigma(e^+e^-\to Z^*H)_{\rm SM}$. For the MSSM the
following relations hold
\begin{eqnarray}
\sigma(e^+e^-\to Z^*h)_{\rm SUSY}&=&\sin^2(\alpha-\beta)\sigma(e^+e^-\to
						Z^*H)_{\rm SM},\\
{\rm BR}(h\to2\,{\rm jets})_{\rm SUSY}&=&f\cdot{\rm BR}(H\to2\,{\rm jets})_{\rm
							SM}.
\end{eqnarray}
{}From Eq.~(\ref{LI}) one can deduce the integrated luminosity achieved,\\
$\int{\cal L}dt=3/(\sigma_{\rm SM}(61.6){\rm BR}_{\rm SM})$. In analogy with
Eq.~(\ref{LI}), we can write
\beq
\#{\rm events}_{\rm\,SUSY}=\sigma_{\rm SUSY}(m_h)\times{\rm BR}_{\rm
SUSY}\times\int{\cal L}dt=3f\cdot\sigma_{\rm SUSY}(m_h)/\sigma_{\rm
SM}(61.6)<3.
\eeq
This immediately implies the following condition for {\it allowed} points in
parameter space \cite{LNPWZh,LG}
\beq
f\cdot\sin^2(\alpha-\beta)<P(61.6/M_Z)/P(m_h/M_Z),
\eeq
where we have used the fact that the cross sections differ simply by the
coupling factor $\sin^2(\alpha-\beta)$ and the Higgs mass dependence which
enters through a function $P$ \cite{HHG}
\beq
P(y)={3y(y^4-8y^2+20)\over\sqrt{4-y^2}}\cos^{-1}
\left(y(3-y^2)\over2\right)-3(y^4-6y^2+4)\ln
y-\coeff{1}{2}(1-y^2)(2y^4-13y+47).
\eeq

The cross section $\sigma_{\rm SUSY}(m_h)$ for the minimal $SU(5)$ model also
corresponds to the SM result since one can verify that
$\sin^2(\alpha-\beta)>0.9999$ in this case. For the flipped model there is a
small deviation ($\sin^2(\alpha-\beta)>0.95$) relative to the SM result for
some points \cite{LNPWZh}. In the calculation of ${\rm BR}(h\to2\,{\rm
jets})_{\rm SUSY}$ which enters in the ratio $f$, we have included {\it all}
contributing modes, in particular the invisible $h\to\chi^0_1\chi^0_1$ decays.
The conclusion is that these models differ little from the SM and in fact the
proper lower bound on $m_h$ is very near $60\GeV$, although it varies from
point to point in the parameter space.

In Ref.~\cite{LNPWZh} it was also shown that this phenomenon is due to a
decoupling effect of the Higgs sector as the supersymmetry scale rises, and
it is communicated to the Higgs sector through the radiative electroweak
symmetry breaking mechanism. The point to be stressed is that if the
supersymmetric Higgs sector is found to be SM-like, this could be taken as
{\em indirect} evidence for an underlying radiative electroweak breaking
mechanism, since no insight could be garnered from the MSSM itself.

Note that since the lower bound on the SM Higgs boson mass could still be
pushed up several GeV at LEPI, the strict dilaton scenario in
Sec.~\ref{specialDilaton} (which requires $m_h\approx61-91\GeV$) could be
further constrained at LEPI.

\begin{figure}[p]
\vspace{4.5in}
%\special{psfile=proc_summer7a.ps angle=90 hscale=60 vscale=60 hoffset=440}
\vspace{-2.5in}
\caption{\baselineskip=12pt
The number of ``mixed" events (1-lepton+2jets+$\mpt$) events per ${\cal
L}=100\ipb$ at LEPII versus the chargino mass in the minimal $SU(5)$ model.}
\label{Figure7a}
\vspace{5in}
%\special{psfile=proc_erice7.ps angle=90 hscale=60 vscale=60 hoffset=440}
\vspace{-0.3in}
\caption{\baselineskip=12pt
The number of ``mixed" events (1-lepton+2jets+$\mpt$) events per ${\cal
L}=100\ipb$ at LEPII versus the chargino mass in the no-scale model (top row).
Also shown (bottom row) are the number of di-electron events per ${\cal
L}=100\ipb$  from selectron pair production versus the lightest selectron
mass.}
\label{Figure7}
\end{figure}

\subsection{LEPII}
\begin{description}
\item (a)  At LEPII the SM Higgs mass could be explored up to roughly the beam
energy minus $100\GeV$ \cite{Alcaraz}. This will allow exploration of almost
al of the Higgs parameter space in the minimal $SU(5)$ model \cite{LNPWZ}.
In the flipped $SU(5)$ models, only low $\tan\beta$ values could be explored,
although the strict no-scale case will probably be out of reach (see Figs.
\ref{Figure4},\ref{Figure5}). The $e^+e^-\to hA$ channel will be open in the
flipped $SU(5)$ models for large $\tan\beta$ and low $m_{\tilde g}$. This
channel is always closed in the minimal $SU(5)$ case (since $m_A\gsim1\TeV$).
It is important to point out that the preferred $h\to b\bar b,c\bar c,gg$
detection modes may be suppressed because of invisible Higgs decays
($h\to\chi^0_1\chi^0_1$) for $m_h\lsim80\GeV$ ($m_h\gsim80\GeV$) by as much
as 30\%/15\% (40\%/40\%) in the minimal/flipped $SU(5)$ model \cite{LNPWZ}.
\item (b) Chargino masses below the kinematical limit
($m_{\chi^\pm_1}\lsim100\GeV$) should not be a problem to detect through the
``mixed" mode with one chargino decaying leptonically and the other one
hadronically \cite{LNPWZ}, \ie, $e^+e^-\to\chi^+_1\chi^-_1$, $\chi^+_1\to
\chi^0_1 q\bar q'$, $\chi^-_1\to\chi^0_1 l^-\bar\nu_l$. In
Fig.~\ref{Figure7a} and Fig.~\ref{Figure7} (top row) we show the correponding
event rates in the minimal $SU(5)$ and no-scale flipped $SU(5)$ models. Recall
that $m_{\chi^\pm_1}$ can be as high as $\approx290\GeV$ in the flipped models,
whereas $m_{\chi^\pm_1}\lsim100\GeV$ in the minimal $SU(5)$ model.
Interestingly enough, the number of mixed events do not overlap (they are much
higher in the minimal $SU(5)$ model) and therefore, if $m_{\chi^\pm_1}<100\GeV$
then LEPII should be able to exclude at least of the models.
\item (c) Selectron, smuon, and stau pair production is partially accessible
for both the no-scale and dilaton models (although more so in the no-scale
case), and completely inaccessible in the minimal $SU(5)$ case. In Fig.
\ref{Figure7} (bottom row) we show the rates for the most promising (dilepton)
mode in $e^+e^-\to\tilde e^+_R\,\tilde e^-_R$ production in the no-scale model.
\end{description}

\subsection{HERA} The elastic and deep-inelastic contributions to
$e^-p\to\tilde e^-_R\chi^0_1$ and $e^-p\to\tilde\nu\chi^-_1$ at HERA in the
no-scale flipped $SU(5)$ model should push the LEPI lower bounds on the
lightest selectron, the lightest neutralino, and the sneutrino masses by
$\approx25\GeV$ with ${\cal L}=100\ipb$ \cite{hera}. In Fig. \ref{Figure8} we
show the elastic plus deep-inelastic contributions to the total supersymmetric
signal ($ep\to{\rm susy}\to eX+\mpt$) versus the lightest selectron mass
($m_{\tilde e_R}$) and the sneutrino mass $(m_{\tilde\nu})$ in the no-scale
model. These figures show the ``reach" of HERA in each of these variables. With
${\cal L}=1000\ipb$ HERA should be competitive with LEPII as far as the
no-scale model is concerned. In the dilaton scenario, because of the somewhat
heavier sparticle masses, the effectiveness of HERA is reduced, although
probably both channels may be accessible. HERA is not sensitive
to the minimal $SU(5)$ model spectrum.

\begin{figure}[t]
\vspace{4.7in}
%\special{psfile=proc_erice8.ps angle=90 hscale=60 vscale=60 hoffset=440}
\vspace{-0.3in}
\caption{\baselineskip=12pt
The elastic plus deep-inelastic total supersymmetric cross section at HERA
($ep\to{\rm susy}\to eX+\mpt$) versus the lightest selectron mass ($m_{\tilde
e_R}$) and the sneutrino mass ($m_{\tilde\nu}$). The short- and long-term
limits of sensitivity are expected to be $10^{-2}\pb$ and $10^{-3}\pb$
respectively.}
\label{Figure8}
\end{figure}

\section{Prospects for Indirect Experimental Detection}
\label{Indirect}
\subsection{$\bf{b\to s\gamma}$}
There has recently been a renewed surge of interest on the
flavor-changing-neutral-current (FCNC) $b\to s\gamma$ decay, prompted by the
CLEO 95\% CL allowed range\cite{Thorndike}
\beq
{\rm BR}(b\to s\gamma)=(0.6-5.4)\times10^{-4}.
\eeq
Since the Standard Model (SM) prediction looms around $(2-5)\times10^{-4}$
depending on the top-quark mass ($m_t$), a reappraisal of beyond the SM
contributions has become topical \cite{Barger,BG,bsgamma,bsgammaII}.  We use
the following expression for the branching ratio $b\to s\gamma$ \cite{BG}
\beq
{{\rm BR}(b\to s\gamma)\over{\rm BR}(b\to ce\bar\nu)}={6\alpha\over\pi}
{\left[\eta^{16/23}A_\gamma
+\coeff{8}{3}(\eta^{14/23}-\eta^{16/23})A_g+C\right]^2\over
I(m_c/m_b)\left[1-\coeff{2}{3\pi}\alpha_s(m_b)f(m_c/m_b)\right]},
\eeq
where $\eta=\alpha_s(M_Z)/\alpha_s(m_b)$, $I$ is the phase-space factor
$I(x)=1-8x^2+8x^6-x^8-24x^4\ln x$, and $f(m_c/m_b)=2.41$ the QCD
correction factor for the semileptonic decay. The $A_\gamma,A_g$ are the
coefficients of the effective $bs\gamma$ and $bsg$ penguin operators
evaluated at the scale $M_Z$. Their simplified expressions are given in
the Appendix of Ref.~\cite{BG}, where the gluino and neutralino contributions
have been justifiably neglected \cite{Bertolini} and the squarks are considered
degenerate in mass, except for the $\tilde t_{1,2}$ which are significantly
split by $m_t$.

For the minimal $SU(5)$ supergravity model we find \cite{bsgamma}
\beq
2.3\,(2.6)\times10^{-4}<{\rm BR}(b\to
s\gamma)_{minimal}<3.6\,(3.3)\times10^{-4},
\eeq
for $\mu>0\,(\mu<0)$, which are all within the CLEO limits. One can show
that ${\rm BR}(b\to s\gamma)$ would need to be measured with better than $20\%$
accuracy to start disentangling the minimal $SU(5)$ supergravity model from the
SM.
In Fig.~\ref{Figure15} we present the analogous results for the no-scale (top
row) and strict no-scale (bottom row) flipped $SU(5)$ supergravity models.
The results are strikingly different than in the prior case.  One observes that
part of the parameter space is actually excluded by the new CLEO bound, for a
range of sparticle masses. Perhaps the most surprising feature of the results
is the strong suppression of ${\rm BR}(b\to s\gamma)$ which occurs for a good
portion of the parameter space for $\mu>0$. It has proven to be non-trivial to
find a simple explanation for the observed cancellation. See
Ref.~\cite{bsgammaII} for the implications of this indirect constraint on the
prospects for direct experimental detection of these models.

\begin{figure}[t]
\vspace{4.7in}
%\special{psfile=proc_summer15.ps angle=90 hscale=60 vscale=60 hoffset=450}
\vspace{-0.3in}
\caption{\baselineskip=12pt
The calculated values of ${\rm BR}(b\to s\gamma)$ versus the chargino mass in
the no-scale (top row) and strict no-scale (bottom row, for the indicated
values of $m_t$) flipped $SU(5)$ supergravity models. Note the fraction of
parameter space excluded by the CLEO allowed range (in between the dashed
lines).}
\label{Figure15}
\end{figure}

\subsection{$\bf{\epsilon_{1,2,3}}$}
A complete study of one-loop electroweak radiative corrections in the
supergravity models we consider here has been made in Ref.~\cite{ewcorr}.
These calculations include some recently discovered important $q^2$-dependent
effects, which occur when light charginos ($m_{\chi^\pm}\lsim 60-70\GeV$) are
present \cite{BFC}, and lead to strong correlations between the chargino and
the top-quark mass. Specifically, one finds that at present the $90\%$ CL upper
limit on the top-quark mass is $m_t\lsim175\GeV$ in the no-scale flipped
$SU(5)$ supergravity model. These bounds can be strengthened for increasing
chargino masses in the $50-100\GeV$ interval. For example, in the flipped model
for $m_{\chi^\pm_1}\gsim60\,(70)\GeV$, one finds $m_t\lsim165\,(160)\GeV$. As
expected, the heavy sector of both models decouples quite rapidly with
increasing sparticle masses, and at present, only $\epsilon_1$ leads to
constraints on the parameter spaces of these models. For future reference, it
is important to note that global SM fits to all of the low-energy and
electroweak data constrain $m_t<194,178,165\GeV$ for
$m_{H_{SM}}=1000,250,50\GeV$ at the $90\%$ CL respectively \cite{PDG}.

One can show that an expansion of the vacuum polarization tensors to order
$q^2$, results in three independent physical parameters. In the first
scheme introduced to study these effects \cite{PT}, namely the $(S,T,U)$
scheme, a SM reference value for $m_t,m_{H_{SM}}$ is used, and the deviation
from this reference is calculated and is considered to be ``new" physics. This
scheme is only valid to lowest order in $q^2$, and is therefore not applicable
to a theory with new, light $(\sim M_Z)$ particles. In the supergravity
models we consider here, each point in parameter space is actually a distinct
model, and a SM reference point is not meaningful. For these reasons, in
Ref.~\cite{ewcorr} the scheme of Refs. \cite{AB,BFC} was chosen, where the
contributions are {\it absolute} and valid to higher order in $q^2$. This
is the so-called $\epsilon_{1,2,3}$ scheme. Regardless of the scheme used, all
of the global fits to the three physical parameters are {\it entirely
consistent} with the SM at $90\%$ CL.

With the assumption that the dominant ``new" contributions arise from the
process-independent (\ie, ``oblique") vacuum polarization amplitudes, one
can combine several observables in suitable ways such that they are most
sensitive to new effects. It is important to note that not all observables
can be included in the experimental fits which determine the $\epsilon_{1,2,3}$
paremeters, if only the oblique contributions are kept \cite{ewcorr}.

It is well known in the MSSM that the largest contributions to $\epsilon_1$
(\ie, $\delta\rho$ if $q^2$-dependent effects are neglected) are expected to
arise from the $\tilde t$-$\tilde b$ sector, and in the limiting case of a very
light stop, the contribution is comparable to that of the $t$-$b$ sector
\cite{DH}. The remaining squark, slepton, chargino, neutralino, and Higgs
sectors all typically contribute considerably less. For increasing sparticle
masses, the heavy sector of the theory decouples, and only SM effects  with a
{\it light} Higgs survive. However, for very light chargino, a $Z$-wavefunction
renormalization threshold effect can introduce a substantial $q^2$-dependence
in the calculation, thus modifying significantly the standard $\delta\rho$
results.  For completeness, in Ref.~\cite{ewcorr} the complete vacuum
polarization contributions from the Higgs sector, the supersymmetric
chargino-neutralino and sfermion sectors, and also the corresponding
contributions in the SM were included.

In Fig.~\ref{Figure16} we show the calculated values of $\epsilon_1$ versus the
lightest chargino mass ($m_{\chi^\pm_1}$) for the sampled points in the minimal
(no-scale flipped) $SU(5)$ supergravity model. In the no-scale flipped $SU(5)$
case three representative values of $m_t$ were used, $m_t=100,130,160\GeV$,
whereas in the minimal $SU(5)$ case several other values for $m_t$ in the range
$90\GeV\le m_t\le 160\GeV$ were sampled. In both models, but most clearly in
the no-scale model one can see how quickly the sparticle spectrum decouples as
$m_{\chi^\pm_1}$ increases, and the value of $\epsilon_1$ asymptotes to the SM
value appropriate to each value of $m_t$ and for a {\it light} ($\sim 100
\GeV$) Higgs mass. The threshold effect of $\chi^\pm_1$ is manifest as
$m_{\chi^\pm_1}\rightarrow {1\over2} M_Z$ and is especially visible for $\mu<0$
in both models. This effect is not expected to be very accurate as
$m_{\chi^\pm_1}\rightarrow {1\over 2} M_Z$. However, according to
Ref.~\cite{BFC}, for $m_{\chi^\pm_1}>50\GeV$, this correction agrees to better
than $10\%$ with the one obtained in a more accurate way.

\begin{figure}[t]
\vspace{4.7in}
%\special{psfile=proc_summer16.ps angle=90 hscale=60 vscale=60 hoffset=450}
\vspace{-0.3in}
\caption{\baselineskip=12pt
The total contribution to $\epsilon_1$ as a function
of the lightest chargino mass $m_{\chi^\pm_1}$ for the minimal $SU(5)$ model
(upper row) and the no-scale flipped $SU(5)$ model (bottom row). Points between
the two horizontal solid lines are allowed at $90\%$ CL. The three distinct
curves (from lowest to highest) in then no-scale case correspond to
$m_t=100,130,160 \GeV$.}
\label{Figure16}
\end{figure}

Recent values for $\epsilon_{1,2,3}$ obtained from a global fit to the
LEP (\ie, $\Gamma_l,A^{l,b}_{FB},A^\tau_{pol})$ and $M_W/M_Z$ measurements
are \cite{BB},
\beq
\epsilon_1=(-0.9\pm
3.7)10^{-3},\quad\epsilon_2=(9.9\pm8.0)10^{-3},\quad\epsilon_3=(-0.9\pm
4.1)10^{-3}.
\eeq
For $\epsilon_1$ it is clear that virtually all the sampled points in the
minimal $SU(5)$ supergravity model are within the $\pm 1.64 \sigma$
($90\%$ CL) bounds (denoted by the two horizontal solid lines in the figures).
Since several values for $90\le m_t\le 160\GeV$ were sampled, the trends for
fixed $m_t$ are not very clear from the figure. Nonetheless, the points
just outside the $1.64 \sigma$ line correspond to $m_t=160\GeV$, which
are therefore excluded at the $90\%$ CL. In the no-scale model, the upper bound
on $m_t$ depends sensitively on the chargino mass. For example, for
$m_t=160\GeV$, only light chargino masses would be acceptable at $90\%$ CL. In
fact,  in Ref.~\cite{ewcorr} the region $130\GeV\le m_t\le 190\GeV$ was
scanned in increments of $5\GeV$ and obtained the maximum values
for $m_{\chi^\pm_1}$ allowed by the experimental value for $\epsilon_1$ at
$90\%$ CL. These are given in the Table~\ref{Table6}. One can immediately see
the strong correlation between $m_t$ and $m_{\chi^\pm_1}$: as $m_t$ rises, the
upper limit to $m_{\chi^\pm_1}$ falls, and vice versa. In particular, for
$m_t\le150\GeV$ all values of $m_{\chi^\pm_1}$ are allowed, while one could
have $m_t$ as large as $160\,(175)\GeV$ for $\mu>0\,(\mu<0)$ if the chargino
mass were light enough.

\begin{table}
\hrule
\caption{
Maximum allowed  chargino mass ($m_{\chi^\pm_1}$) for different $m_{t}$ (in
GeV) at $90\% $CL in the no-scale flipped $SU(5)$ model. In the entries Y(N)
means all points are within (outside) the LEP bounds at $90\% $CL.}
\label{Table6}
\begin{center}
\begin{tabular}{|c|c|c|}\hline
$m_{t}$&$\mu>0$&$\mu<0$\\ \hline
145& Y & Y   \\
150& Y & Y   \\
155&68& 95 \\
160&66& 72 \\
165& N & 63 \\
170& N & 58 \\
175& N & 53 \\
180& N & N   \\ \hline
\end{tabular}
\end{center}
\hrule
\end{table}

%r \subsection{Neutrino telescopes}
%r \subsection{$(g-2)_{\mu,\tau}$}

\section{Conclusions}
\label{Conclusions}
The recent surge of interest in supersymmetric models, spurred by the precise
LEP measurements of the gauge couplings and their unification at very high
energies \cite{EKNI}, has made it clear that some sort of organizing principle
is needed to tame the zoo of supersymmetric particles at low energies, as
encompassed by the minimal supersymmetric standard model (MSSM). Even though it
is usually not acknowledged that this model needs at least {\em twenty-one
parameters} for its full description, the fact remains that this is the case.
One reason for disregarding this fact is that many phenomenological
calculations of interest have been performed and numerical results obtained by
making ad-hoc assumptions about the values of these parameters, giving the
erroneous impression that the results so-obtained are largely insensitive to
changes in these assumptions. This mindset has been explicitly exposed by now
on a variety of calculations. These have been performed in the context of the
class of unified supergravity models which we consider here, and have been
found to yield quite different results than previously expected. A few examples
include: the calculation of the cosmological relic density of neutralinos in
the MSSM \cite{minimaldm} and in supergravity models without
\cite{without,LNYdmI,LNYdmII} and with \cite{with,KLNPYdm} radiative
electroweak symmetry breaking; the enhancement of the leptonic branching ratios
of the chargino in the presence of light sleptons \cite{LNWZ}; the large
sparticle mass dependence of the $b\to s\gamma$ branching ratio \cite{bsgamma}
which can completely wash out the charged Higgs contribution \cite{Barger};
etc. It then becomes apparent that one must make further well motivated
theoretical assumptions to make further headway into this problem. As discussed
above, supergravity models with radiative electroweak symmetry breaking
accomplish this goal very economically, needing just three supersymmetry
breaking parameters, the ratio $\tan\beta$, and the soon-to-be-measured
top-quark mass to be fully described. We have discussed two classes of such
models, the class of string-inspired flipped $SU(5)$ models and the traditional
minimal $SU(5)$ supergravity model.

The simplest, string-derivable, supergravity model has as gauge group flipped
$SU(5)$ with supplementary matter representations to ensure unification at the
string scale ($\sim10^{18}\GeV$). This basic structure is complemented by two
possible string supersymmetry breaking scenaria: $SU(N,1)$ no-scale
supergravity and dilaton-induced supersymmetry breaking. These two variants
should be considered to be idealizations of what their string-derived
incarnation should be. The specification of the hidden sector is crucial to the
determination of the supersymmetry breaking scenario at work. A thorough
exploration of the parameter spaces of the two models yields interesting
results for direct and indirect experimental detection at present or near
future colliders. In this regard, the no-scale model is more within reach than
the dilaton model, because of its generally lighter spectrum. In both
supersymmetry breaking scenaria considered, there ia a more constrained special
case which allows $\tan\beta$ to be determined in terms of $m_t$ and $m_{\tilde
g}$. In the strict no-scale case we find a striking result: if $\mu>0$,
$m_t\lsim135\GeV$, whereas if $\mu<0$, $m_t\gsim140\GeV$. Therefore the value
of $m_t$ determines the sign of $\mu$. Furthermore, we found that the value of
$m_t$ also determines  whether the lightest Higgs boson is above or below
$100\GeV$. In the restricted dilaton case there is an upper bound on the
top-quark mass ($m_t\lsim155\GeV$) and the lightest Higgs boson mass
($m_h\lsim91\GeV$). Thus, continuing Tevatron top-quark searches and LEPI,II
Higgs searches could probe this restricted scenario completely. In Table
\ref{Table4} we give a summary of the general properties of these models and a
comparison of their spectra.

The minimal $SU(5)$ supergravity model is strongly constrained by the proton
lifetime and the cosmological neutralino relic density. The former constraint
implies light chargino masses and heavy scalar masses (beyond the reach of
present or near future colliders), while the second constraint nearly
enforces the mass relations $m_\chi\approx{1\over2}m_{h,Z}$. The prospects
for detection of the light particles in this model
($\chi^0_{1,2},\chi^\pm_1,h$) is very promising in the near future.
It is interesting to remark that the resulting allowed choices for the
supersymmetry breaking parameters bear close resemblance to those predicted in
moduli-induced string-inspired supersymmetry breaking scenaria. In Table
\ref{Table7} we give a summary of the general properties of this models and its
spectrum.

\begin{table}[p]
\hrule
\caption{Major features of the $SU(5)\times U(1)$ string-inspired/derived model
and a comparison of the two supersymmetry breaking scenaria considered.
(All masses in GeV).}
\label{Table4}
\begin{center}
\begin{tabular}{|l|}\hline
\hfil$\bf SU(5)\times U(1)$\hfil\\ \hline
$\bullet$ Easily string-derivable, several known examples\\
$\bullet$ Symmetry breaking to Standard Model due to vevs of \r{10},\rb{10}\\
\quad and tied to onset of supersymmetry breaking\\
$\bullet$ Natural doublet-triplet splitting mechanism\\
$\bullet$ Proton decay: $d=5$ operators very small\\
$\bullet$ Baryon asymmetry through lepton number asymmetry\\
\quad (induced by the decay of flipped neutrinos) as processed by\\
\quad non-perturbative electroweak interactions\\
\hline
\end{tabular}
\end{center}
\begin{center}
\begin{tabular}{|l|l|}\hline
\hfil$\vev{F_M}_{m_0=0}$ (no-scale)\hfil&\hfil$\vev{F_D}$ (dilaton)\hfil\\
									\hline
$\bullet$ Parameters 3: $m_{1/2},\tan\beta,m_t$&
$\bullet$ Parameters 3: $m_{1/2},\tan\beta,m_t$\\
$\bullet$ Universal soft-supersymmetry&$\bullet$ Universal soft-supersymmetry\\
\quad breaking automatic&\quad breaking automatic\\
$\bullet$ $m_0=0$, $A=0$&$\bullet$ $m_0=\coeff{1}{\sqrt{3}}m_{1/2}$,
$A=-m_{1/2}$\\
$\bullet$ Dark matter: $\Omega_\chi h^2_0<0.25$
&$\bullet$ Dark matter: $\Omega_\chi h^2_0<0.90$\\
$\bullet$ $m_{1/2}<475\GeV$, $\tan\beta<32$
&$\bullet$ $m_{1/2}<465\GeV$, $\tan\beta<46$\\
$\bullet$ $m_{\tilde g}>245\GeV$, $m_{\tilde q}>240\GeV$
&$\bullet$ $m_{\tilde g}>195\GeV$, $m_{\tilde q}>195\GeV$\\
$\bullet$ $m_{\tilde q}\approx0.97m_{\tilde g}$
&$\bullet$ $m_{\tilde q}\approx1.01m_{\tilde g}$\\
$\bullet$ $m_{\tilde t_1}>155\GeV$
&$\bullet$ $m_{\tilde t_1}>90\GeV$\\
$\bullet$ $m_{\tilde e_R}\approx0.18m_{\tilde g}$,
$m_{\tilde e_L}\approx0.30m_{\tilde g}$
&$\bullet$ $m_{\tilde e_R}\approx0.33m_{\tilde g}$,
$m_{\tilde e_L}\approx0.41m_{\tilde g}$\\
\quad $m_{\tilde e_R}/m_{\tilde e_L}\approx0.61$
&\quad $m_{\tilde e_R}/m_{\tilde e_L}\approx0.81$\\
$\bullet$ $60\GeV<m_h<125\GeV$&$\bullet$ $60\GeV<m_h<125\GeV$\\
$\bullet$ $2m_{\chi^0_1}\approx m_{\chi^0_2}\approx m_{\chi^\pm_1}\approx
0.28 m_{\tilde g}\lsim290$
&$\bullet$ $2m_{\chi^0_1}\approx m_{\chi^0_2}\approx m_{\chi^\pm_1}\approx
0.28 m_{\tilde g}\lsim285$\\
$\bullet$ $m_{\chi^0_3}\sim m_{\chi^0_4}\sim m_{\chi^\pm_2}\sim\vert\mu\vert$
&$\bullet$ $m_{\chi^0_3}\sim m_{\chi^0_4}\sim m_{\chi^\pm_2}\sim\vert\mu\vert$
\\
$\bullet$ Spectrum easily accessible soon
&$\bullet$ Spectrum accessible soon\\ \hline
$\bullet$ Strict no-scale: $B(M_U)=0$
&$\bullet$ Special dilaton: $B(M_U)=2m_0$\\
\quad $\tan\beta=\tan\beta(m_t,m_{\tilde g})$
&\quad $\tan\beta=\tan\beta(m_t,m_{\tilde g})$\\
\quad $m_t\lsim135\GeV\Rightarrow\mu>0,m_h\lsim100\GeV$
&\quad $\tan\beta\approx1.4-1.6$, $m_t<155\GeV$\\
\quad $m_t\gsim140\GeV\Rightarrow\mu<0,m_h\gsim100\GeV$
&\quad $m_h\approx61-91\GeV$\\
\hline
\end{tabular}
\end{center}
\hrule
\end{table}

\begin{table}[t]
\hrule
\caption{Major features of the minimal $SU(5)$ supergravity model and its
spectrum (All masses in GeV).}
\label{Table7}
\begin{center}
\begin{tabular}{|l|}\hline
\hfil$\bf SU(5)$\hfil\\ \hline
$\bullet$ Not easily string-derivable, no known examples\\
$\bullet$ Symmetry breaking to Standard Model due to vevs of \r{24}\\
\quad and independent of supersymmetry breaking\\
$\bullet$ No simple doublet-triplet splitting mechanism\\
$\bullet$ Proton decay: $d=5$ operators large, strong constraints needed\\
$\bullet$ Baryon asymmetry ?\\ \hline
\end{tabular}
\end{center}
\begin{center}
\begin{tabular}{|l|l|}\hline
\hfil Spectrum\hfil\\ 	\hline
$\bullet$ Parameters 5: $m_{1/2},m_0,A,\tan\beta,m_t$\\
$\bullet$ Universal soft-supersymmetry breaking automatic\\
$\bullet$ $m_0/m_{1/2}>3$, $\tan\beta\lsim3.5$\\
$\bullet$ Dark matter: $\Omega_\chi h^2_0\gg1$, 1/6 of points excluded\\
$\bullet$ $m_{\tilde g}<400\GeV$,
$m_{\tilde q}>m_{\tilde l}>2m_{\tilde g}\gsim500\GeV$\\
$\bullet$ $m_{\tilde t_1}>45\GeV$\\
$\bullet$ $60\GeV<m_h<100\GeV$\\
$\bullet$ $2m_{\chi^0_1}\approx m_{\chi^0_2}\approx m_{\chi^\pm_1}\approx
0.28 m_{\tilde g}\lsim100$\\
$\bullet$ $m_{\chi^0_3}\sim m_{\chi^0_4}\sim m_{\chi^\pm_2}\sim\vert\mu\vert$
\\
$\bullet$ Chargino and Higgs easily accessible soon\\
\hline
\end{tabular}
\end{center}
\hrule
\end{table}

We conclude that these well motivated supergravity models (especially the
strict versions of the string-inspired/derived models) could soon be probed
experimentally. The various ingredients making up the flipped $SU(5)$ models
are likely to be present in actual fully string-derived models which yield the
set of supersymmetry breaking parameters in Eqs. (\ref{nsc},\ref{kl}). The
search for such models is imperative, although it may not be an easy task since
in traditional gaugino condensation scenaria Eqs. (\ref{nsc},\ref{kl}) are
usually not reproduced (see however Refs. \cite{Ross,Zwirner}). Moreover, the
requirement of vanishing vacuum energy may be difficult to fulfill, as a model
with these properties and all the other ones outlined in Sec.~\ref{Typical} is
yet to be found. This should not be taken as a discouragement since the harder
it is to find the correct model, the more likely it is to be in some sense
unique.

\section*{Acknowledgements}
This work has been supported in part by DOE grant DE-FG05-91-ER-40633. The work
of J.L. has been supported by an SSC Fellowship.

%r \appendix

\end{document}